\documentclass[11pt]{article}%
\usepackage{amsmath}
\usepackage{amssymb}
\usepackage{amsfonts}

\usepackage{cite}
\usepackage{graphicx}
\usepackage{float}
\usepackage{longtable}
\usepackage[utf8]{inputenc}
\usepackage{hyperref}

\usepackage{algorithmicx}
\usepackage[ruled,vlined]{algorithm2e}
\SetAlgoCaptionSeparator{ }

\usepackage{fullpage}

\newcommand{\fref}[1]{Fig.~\ref{#1}}

\newcommand{\tref}[1]{Table~\ref{#1}}
\newcommand{\sref}[1]{Section~\ref{#1}}

\newenvironment{method}[1][!htbp]
  {
   \begin{algorithm}[#1]%
  }{\end{algorithm}}

\newenvironment{subproc}[1][h!]
  {
   \begin{algorithm}[#1]%
  }{\end{algorithm}}

\providecommand{\U}[1]{\protect\rule{.1in}{.1in}}

\newtheorem{problem}{Problem}

\begin{document}
\title{A Poisson Model for Entanglement Optimization in the Quantum Internet}
\author{Laszlo Gyongyosi\thanks{School of Electronics and Computer Science, University of Southampton, Southampton SO17 1BJ, U.K., and Department of Networked Systems and Services, Budapest University of Technology and Economics, 1117 Budapest, Hungary, and MTA-BME Information Systems Research Group, Hungarian Academy of Sciences, 1051 Budapest, Hungary.}
\thanks{Parts of this work were presented in conference proceedings \cite{ref8}.}
\and Sandor Imre\thanks{Department of Networked Systems and Services, Budapest University of Technology and Economics, 1117 Budapest, Hungary.}}

\date{}

\maketitle
\vspace{-0.5cm}
\begin{abstract}
We define a nature-inspired model for entanglement optimization in the quantum Internet. The optimization model aims to maximize the entanglement fidelity and relative entropy of entanglement for the entangled connections of the entangled network structure of the quantum Internet. The cost functions are subject of a minimization defined to cover and integrate the physical attributes of entanglement transmission, purification, and storage of entanglement in quantum memories. The method can be implemented with low complexity that allows a straightforward application in the quantum Internet and quantum networking scenarios.
\end{abstract}

\section{Introduction}
\label{sec1}
Quantum entanglement and the entangled network structure serve as fundamental concepts of the quantum Internet \cite{ref1,ref3,ref4,ref5,ref6}, long-distance quantum networks and future quantum communications \cite{ref29,ref30,ref31,ref32,ref33,ref34,ref35,ref36,ref37,ref38,ref39,ref40,ref41,ref42,ref43,ref44}. Since the no-cloning theorem makes it impossible to use the ``copy-and-resend'' mechanisms of traditional repeaters \cite{ref7,ref45}, in a quantum Internet scenario the quantum repeaters have to transmit correlations in a different way \cite{ref1,ref2,ref3,ref4,ref5}. In the entangled network structure of the quantum Internet, the main task of quantum repeaters is to distribute quantum entanglement between distant points that will then serve as a fundamental base resource for quantum teleportation and other quantum protocols \cite{ref1}. Since in an experimental scenario \cite{ref15, ref16, ref17, ref18, ref19, ref20, ref21} the quantum links between nodes are noisy and entanglement fidelity decreases as hop distance increases, entanglement purification is applied to improve the entanglement fidelity between nodes \cite{ref1, ref3, ref4, ref5, ref6}. Quantum nodes also perform internal quantum error correction that is a requirement for reliability and storage in quantum memories \cite{ref1, ref5, ref6, ref8, ref50,ref51,ref52,ref53,ref54,ref55,ref56,ref57,ref58,ref59,ref60}. Both entanglement purification and quantum error correction steps in local nodes are high-cost tasks that require significant minimization \cite{ref1, ref3, ref4, ref5, ref6, ref15, ref16, ref19, ref20, ref21, ref22, ref23, ref24, ref25, ref26, ref27, ref28, ref29, ref30, ref31, ref32, ref47,ref48,ref49,ref61,ref62,ref63,ref64,ref65,ref66,ref67,ref68,ref69,ref70,ref71,ref72,ref73,ref74,ref75,ref76,ref77,ref78,ref79,ref80,ref81,ref82,ref83,ref84}. 

The shared entangled states between nodes form entangled connections. Significant attributes of these entangled connections are entanglement fidelity \cite{ref1,ref5,ref6}, and correlation in terms of relative entropy entanglement \cite{ref76,ref77}. Entanglement fidelity is a crucial parameter. It serves as the primary objective function in our model, which is a subject of maximization. Maximizing the relative entropy of entanglement is the secondary objective function. Minimizing the cost of classical communications, which is required by the entanglement optimization method as an auxiliary objective function, is also considered.

Besides these attributes, the entangled connections are characterized by the entanglement throughput that identifies the number of transmittable entangled systems per sec at a particular fidelity. In our model, the nodes are associated with an incoming entanglement throughput \cite{ref1}, that serves as a resource for the nodes to maximize the entanglement fidelity and the relative entropy of entanglement. The nodes receive and process the incoming entangled states. Each node performs purification and internal quantum error correction, and it stores the entangled systems in local quantum memories. The amount of input entangled systems in a node is therefore connected to the achievable maximal entanglement fidelity and correlation in the entangled states associated with that node. The objective of the proposed model is to reveal this connection and to define a framework for entanglement optimization in the quantum nodes of an arbitrary quantum network. The required input information for the optimization without loss of generality are the number of nodes, the number of fidelity types of the received entangled states, and the node characteristics. In a realistic setting, these cover the incoming entanglement throughput in a node and the costs of internal entanglement purification steps, internal quantum error corrections, and quantum memory usage. 

In this work, an optimization framework for quantum networks is defined. The method aims to maximize the achievable entanglement fidelity and correlation of entangled systems, in parallel with the minimization of the cost of entanglement purification and quantum error correction steps in the quantum nodes of the network. The problem model is therefore defined as a multiobjective optimization. This paper aims to provide a model that utilizes the realistic parameters of the internal mechanisms of the nodes and the physical attributes of entanglement transmission. The proposed framework integrates the results of quantum Shannon theory, the theory of evolutionary multiobjective optimization algorithms \cite{ref9, ref10}, and the mathematical modeling of seismic wave propagation \cite{ref9, ref10, ref11, ref12, ref13, ref14}. 

Inspired by the statistical distribution of seismic events and the modeling of wave propagations in nature, the model utilizes a Poisson distribution framework to find optimal solutions in the objective space. In the theory of earthquake analysis and spatial connection theory \cite{ref9, ref10, ref11, ref12, ref13, ref14}, Poisson distributions are crucial in finding new epicenters. Motivated by these findings, a Poisson model is proposed to find new solutions in the objective space that is defined by the multiobjective optimization problem. The solutions in the objective space are represented by epicenters with several locations around them that also represent solutions in the feasible space \cite{ref9, ref10}. The epicenters have a magnitude and seismic power operators that determine the distributions of the locations and fitness \cite{ref9, ref10} of locations around the epicenters. Epicenters with low magnitude generate high seismic power in the locations, whereas epicenters with high magnitude generate low seismic power in the locations. Epicenters are generated randomly in the feasible space, and each epicenter is weighted from which the magnitude and power are derived. By a general assumption, epicenters with lower magnitude produce more locations because the locations are closer to the epicenter. The locations are placed within a certain magnitude around the epicenters in the feasible space. The optimization framework involves a set of solutions to the Pareto optimal front \cite{ref9,ref10} by combining the concept of Pareto dominance and seismic wave propagations. The new epicenters are determined by a Poisson distribution in analogue to prediction theory in earthquake models. The mathematical model of epicenters allows us to find new solutions iteratively and to find a global optimum. The framework has low complexity that allows an efficient practical implementation to solve the defined multiobjective optimization problem.  

The multiobjective optimization problem model considers the fidelity and correlation of entanglement of entangled states available in the quantum nodes. The resources for the nodes are the incoming entangled states from the quantum links, and the already stored entangled quantum systems in the local quantum memories. In the optimization procedure, both memory consumptions and environmental effects, such as entanglement purification and quantum error correction steps, are considered to develop the cost functions. In particular, the amount of resource, in terms of number of available entangled systems, is a coefficient that can be improved by increasing the incoming number of entangled systems, such as the incoming entanglement throughput in a node. In the proposed model, the incoming entanglement fidelity is further divided into some classes, which allows us to differentiate the resources in the nodes with respect to their fidelity types. Therefore, the fidelity type serves as a quality index for the optimization procedure. The optimization aims to find the optimal incoming entanglement throughput for all nodes that leads to a maximization of entanglement fidelity and correlation of entangled states with respect to the relative entropy of entanglement, for all entangled connections in the quantum network.  

The novel contributions of our manuscript are as follows: 
\begin{enumerate}
\item  \textit{A nature-inspired, multiobjective optimization framework is conceived for the quantum Internet.}
\item \textit{The model considers the physical attributes of entanglement transmission and quantum memories to provide a realistic setting (realistic objective functions and cost functions).}
\item \textit{The method fuses the results of quantum Shannon theory and theory of evolutionary multiobjective optimization algorithms.}
\item \textit{The model maximizes the entanglement fidelity and relative entropy of entanglement for all entangled connections of the network. It minimizes the cost functions to reduce the costs of entanglement purification, error correction, and quantum memory usage.}
\item \textit{The optimization framework allows a low-complexity implementation.}
\end{enumerate}

This paper is organized as follows. \sref{sec2} presents the problem statement. \sref{sec3} details the optimization method. \sref{sec4} provides the problem resolution. \sref{sec5} proposes numerical evidence. Finally, \sref{sec6} concludes the paper. Supplemental material is included in the Appendix. 
 
\section{Problem Statement}
\label{sec2}
The problem to be solved is summarized in Problem 1. 

\begin{problem}
For a given quantum network with $N$ nodes, for all nodes $x_{i} $, $i=1,\ldots ,N$, the entanglement fidelity and relative entropy of entanglement for all entangled connections are maximized, and the cost of optimal purification and quantum error correction and the cost of memory usage for all nodes are minimized. 
\end{problem}

 The network model is as follows. Let $B_{F} \left(x\right)$ be the incoming number of received entangled states (incoming entanglement throughput) in a given quantum node $x$, measured in the number of $d$-dimensional entangled states per sec at a particular entanglement fidelity $F$ \cite{ref1,ref3,ref4}.   

 Let $N$ be the number of nodes in the network, and let $T$ be the number of fidelity types $F_{j} $, $j=1,\ldots ,T$ of the entangled states in the quantum network. 

 Let $B_{F}^{j} \left(x_{i} \right)$ be the number of incoming entangled states in an $i^{th} $ node $x_{i} $, $i=1,\ldots ,N$, from fidelity type $j$. In our model, $B_{F}^{j} \left(x_{i} \right)$ represents the utilizable resources in a particular node $x_{i} $. Thus, the task is to determine this value for all nodes in the quantum network to maximize the fidelity and relative entropy of shared entanglement for all entangled connections. 

 Let $\mathbf{X}$ be an $N\times T$ matrix 
\begin{equation} \label{ZEqnNum640853} 
\mathbf{X}=\left(B_{F}^{j} \left(x_{i} \right)\right)_{N\times T} .                                                 
\end{equation} 
The matrix describes the number of entangled states of each fidelity type for all nodes in the network, $B_{F}^{j} \left(x_{i} \right)\ge 0$ for all $i$ and $j$. 
 
\subsection{Objective Functions}
For a given node $x_{i} $, let ${\rm {\mathcal F}}\left(x_{i} \right)$ be the primary objective function that identifies the cumulative entanglement fidelity (a sum of entanglement fidelities in $x_{i} $) after an entanglement purification ${\rm P} \left(x_{i} \right)$ and an optimal quantum error correction ${\rm C}\left(x_{i} \right)$ in $x_{i} $. In our framework, ${{\mathcal{F}}_{i}}\left( \mathbf{X} \right)$ for a node $x_{i} $ is defined as
\begin{equation} \label{ZEqnNum305166} 
{{\mathcal{F}}_{i}}\left( \mathbf{X} \right)=\sum _{j=1}^{T}\sum _{k=1}^{T}A_{ijk}   \tilde{B}_{F}^{j} \left(x_{i} \right)\tilde{B}_{F}^{k} \left(x_{i} \right)+\sum _{j=1}^{T}R_{ij}  \tilde{B}_{F}^{j} \left(x_{i} \right)+c_{i} ,                          
\end{equation} 
where $A_{ijk} $ is the quadratic regression coefficient, $R_{ij} $ is the simple regression coefficient, $c_{i} $ is a constant, and $\tilde{B}_{F}^{j} \left(x_{i} \right)$ is defined as 
\begin{equation} \label{3)} 
\tilde{B}_{F}^{j}\left( {{x}_{i}} \right)=B_{F}^{j}\left( {{x}_{i}} \right)+\left\langle B \right\rangle _{F}^{j}\left( {{x}_{i}} \right),
\end{equation} 
where $\left\langle B \right\rangle _{F}^{j}\left( {{x}_{i}} \right)$ is an initialization value for $B_{F}^{j} \left(x_{i} \right)$ in a particular node $x_{i} $.

 Then let $\mathbb{E}\left( {{D}_{i}}\left( \mathbf{X} \right) \right)$ be the secondary objective function that refers to the expected amount of cumulative relative entropy of entanglement (a sum of relative entropy of entanglement) in node $x_{i} $, defined as
\begin{equation} \label{ZEqnNum288106} 
\begin{split}
\mathbb{E}\left( {{D}_{i}}\left( \mathbf{X} \right) \right)=&\sum _{j=1}^{T}\sum _{k=1}^{T}A_{ijk}^{*}   \tilde{B}_{F}^{j} \left(x_{i} \right)\tilde{B}_{F}^{k} \left(x_{i} \right)\\&+\sum _{j=1}^{T}R_{ij}^{*}  \tilde{B}_{F}^{j} \left(x_{i} \right)+c_{i}^{*} ,                         
\end{split}
\end{equation} 
where $A_{ijk}^{*} $, $R_{ij}^{*} $, and $c_{i}^{*} $ are some regression coefficients, by definition. 

 Therefore, the aim is to find the values of $B_{F}^{j} \left(x_{i} \right)$ for all $i$ and $j$ in \eqref{ZEqnNum640853}, such that ${{\mathcal{F}}_{i}}\left( \mathbf{X} \right)$ and $\mathbb{E}\left( {{D}_{i}}\left( \mathbf{X} \right) \right)$ are maximized for all $i$. 

 Assuming that the fidelity of entanglement is dynamically changing and evolves over time, the $w_{j} \left(x_{i} \right)$ quantum memory coefficient is introduced for the storage of entangled states from the $j^{th} $ fidelity type in a node $x_{i} $ as follows:
\begin{equation} \label{5)}  
w_{j} \left(x_{i} \right)=\eta _{j} B_{F}^{j} \left(x_{i} \right)+\kappa _{j} \left\langle B \right\rangle _{F}^{j}\left( {{x}_{i}} \right),                                         
\end{equation} 
where $\eta _{j} $ and $\kappa _{j} $ are coefficients that describe the storage characteristic of entangled states with the $j^{th} $ fidelity type.

\subsection{Cost Functions}
The cumulative entanglement fidelity \eqref{ZEqnNum305166} and cumulative relative entropy of entanglement \eqref{ZEqnNum288106} in a particular node $x_{i} $ are associated with a $f_{C} \left({\rm P} \left(x_{i} \right)\right)$ cost entanglement purification ${\rm P} \left(x_{i} \right)$ and a $f_{C} \left({\rm C}\left(x_{i} \right)\right)$ cost of optimal quantum error correction ${\rm C}\left(x_{i} \right)$ in $x_{i} $, where $f_{C} \left(\cdot \right)$ is the cost function. 

Then let $\mathcal{C}\left( \mathbf{X} \right)$ be the total cost function for all of the $T$ fidelity types and for all of the $N$ nodes, as follows:
\begin{equation} \label{ZEqnNum592823} 
\begin{split}
\mathcal{C}\left( \mathbf{X} \right)&=\sum\limits_{i=1}^{N}{{{f}_{C}}\left(\rm P \left( {{x}_{i}} \right) \right)+{{f}_{C}}\left( \text{C}\left( {{x}_{i}} \right) \right)}\\&=\sum\limits_{i=1}^{N}{\sum\limits_{i=1}^{T}{{{f}_{j}}B_{F}^{j}\left( {{x}_{i}} \right),}}
\end{split}
\end{equation} 
where $f_{j} $ is a total cost of purification and error correction associated with the $j^{th} $ fidelity type of entangled states.

Let $F^{{\rm *}} $ be a critical fidelity on the received quantum states. The entangled states are then decomposable into two sets ${\rm {\mathcal S}}_{low} $ and ${\rm {\mathcal S}}_{high} $ with fidelity bounds ${\rm {\mathcal S}}_{low} \left(F\right)$ and ${\rm {\mathcal S}}_{high} \left(F\right)$ as
\begin{equation} \label{ZEqnNum766550} 
{\rm {\mathcal S}}_{low} \left(F\right):\mathop{\max }\limits_{\forall i} F_{i} <F^{{\rm *}} ,                                              
\end{equation} 
and 
\begin{equation} \label{ZEqnNum229590} 
{\rm {\mathcal S}}_{high} \left(F\right):\mathop{\min }\limits_{\forall i} F_{i} \ge F^{{\rm *}} .                                              
\end{equation} 
For the quantum systems of ${\rm {\mathcal S}}_{low} $, the highest fidelity is below the critical amount $F^{{\rm *}} $, and for set ${\rm {\mathcal S}}_{high} $, the lowest fidelity is at least $F^{{\rm *}} $. Then let $X_{{\rm {\mathcal S}}_{low} } $ and $X_{{\rm {\mathcal S}}_{high} } $ identify the set of nodes for which condition \eqref{ZEqnNum766550} or \eqref{ZEqnNum229590} holds, respectively.

Let ${{\mathcal{S}}_{i}}\left( \mathbf{X} \right)$ be the cost of quantum memory usage in node $x_{i} $, defined as 
\begin{equation} \label{ZEqnNum877389} 
{{\mathcal{S}}_{i}}\left( \mathbf{X} \right)=\lambda \sum _{j=1}^{T}\alpha _{i} \frac{1}{\Upsilon _{i} }  B_{F}^{j} \left(x_{i} \right),                                            
\end{equation} 
where $\lambda $ is a constant, $\alpha _{i} $ is a quality coefficient that identifies set \eqref{ZEqnNum766550} or \eqref{ZEqnNum229590} for a given node $x_{i} $, and $\Upsilon _{i} $ is the capacity coefficient of the quantum memory.

\begin{center}
\begin{figure}[!h]
\begin{center}
\includegraphics[angle = 0,width=1\linewidth]{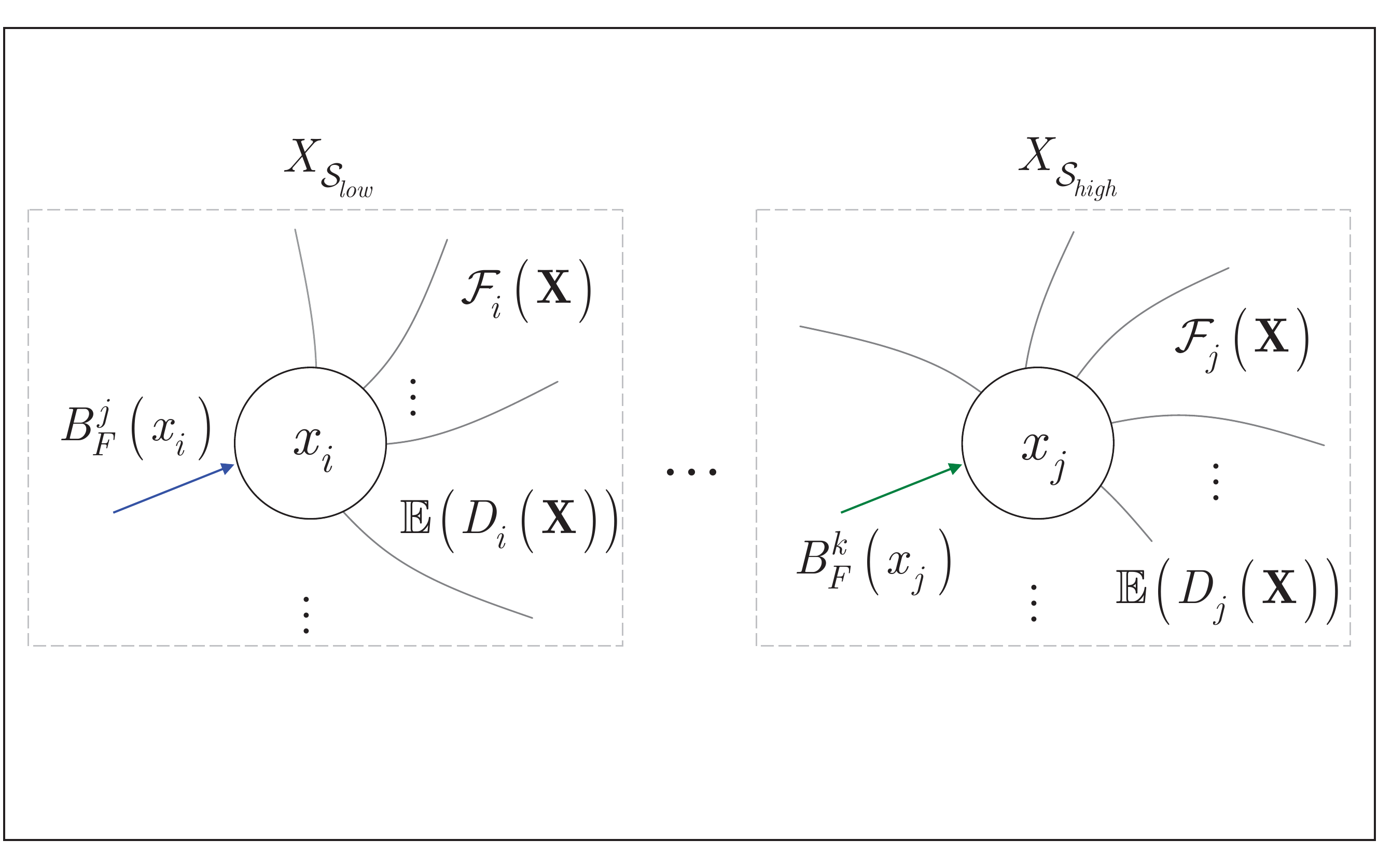}
\caption{Illustration of the network model components. The quantum nodes $x_{i} $ and $x_{j} $ are associated with current input values $B_{F}^{j} \left(x_{i} \right)$ and $B_{F}^{l} \left(x_{j} \right)$ (blue and green arrows), where $j$ and $l$ identify the fidelity types of received entangled states. The nodes have several entangled connections (depicted by gray lines) in the network. The nodes are associated with subject functions ${{\mathcal{F}}_{i}}\left( \mathbf{X} \right)$, $\mathbb{E}\left( {{D}_{i}}\left( \mathbf{X} \right) \right)$, and ${{\mathcal{F}}_{j}}\left( \mathbf{X} \right)$, $\mathbb{E}\left( {{D}_{j}}\left( \mathbf{X} \right) \right)$. The maximum of the received entanglement fidelity in the nodes allows the classification of the nodes to sets $X_{{\rm {\mathcal S}}_{low} } $ and $X_{{\rm {\mathcal S}}_{high} } $: node $x_{i} $ belongs to set $X_{{\rm {\mathcal S}}_{low} } $, whereas node $x_{j} $ belongs to set $X_{{\rm {\mathcal S}}_{high} } $ (depicted by dashed frames).} 
 \label{fig1}
 \end{center}
\end{figure}
\end{center}

 The main components of the network model are depicted in \fref{fig1}. 

\subsection{Multiobjective Optimization}
The optimization problem is as follows. The entanglement fidelity and the relative entropy of entanglement for all types of fidelity of stored entanglement for all nodes are maximized, while the cost of entanglement purification and quantum error correction is minimal, and the memory usage cost (required storage time) is also minimal. These requirements define a multiobjective optimization problem \cite{ref9, ref10}.

Utilizing functions \eqref{ZEqnNum305166} and \eqref{ZEqnNum288106}, the function subject of a maximization to yield maximal entanglement fidelity and maximal relative entropy of entanglement in all nodes of the network is defined via main objective function $\mathcal{G}\left( \mathbf{X} \right)$:
\begin{equation} \label{ZEqnNum737898} 
\mathcal{G}\left( \mathbf{X} \right)=\max \sum _{i=1}^{N}{{\mathcal{F}}_{i}}\left( \mathbf{X} \right)\mathbb{E}\left( {{D}_{i}}\left( \mathbf{X} \right) \right) .                                        
\end{equation} 
Function $\mathcal{G}\left( \mathbf{X} \right)$ should be maximized while cost functions \eqref{ZEqnNum592823} and \eqref{ZEqnNum877389} are minimized via functions $F_{1} \left(N\right)$ and $F_{2} \left(N\right)$:
\begin{equation} \label{11)} 
F_{1} \left(N\right)=\min \mathcal{C}\left( \mathbf{X} \right)=\sum _{i=1}^{N}\sum _{i=1}^{T}f_{j} B_{F}^{j} \left(x_{i} \right)  ,                                     
\end{equation} 
and
\begin{equation} \label{12)} 
{{F}_{2}}\left( N \right)=\min \mathcal{S}\left( \mathbf{X} \right)=\sum\limits_{i=1}^{N}{{{\mathcal{S}}_{i}}\left( \mathbf{X} \right)},                                        
\end{equation} 
with the problem constraints \cite{ref9, ref10} $C_{1} $, $C_{2} $, and $C_{3} $ for all $i$ and $j$. Constraint $C_{1} $ is defined as
\begin{equation} \label{ZEqnNum992191} 
C_{1} :\zeta \left( \mathbf{X} \right)\ge \gamma ,                                                      
\end{equation} 
where $\gamma $ is a cumulative lower bound on the required entanglement fidelity for all nodes, while $\zeta \left( \mathbf{X} \right)$ is 
\begin{equation} \label{14)} 
\zeta \left( \mathbf{X} \right)=\sum _{i=1}^{N}{{\mathcal{F}}_{i}}\left( \mathbf{X} \right) ,                                                   
\end{equation} 
and constraint $C_{2} $ is
\begin{equation} \label{ZEqnNum104266} 
C_{2} :{\rm X} \le \Lambda ,                                                      
\end{equation} 
where $\Lambda $ is an upper bound on the total cost function $\mathcal{C}\left( \mathbf{X} \right)$, while ${\rm X}$ is
\begin{equation} \label{16)} 
{\rm X} =\sum _{i=1}^{N}\sum _{i=1}^{T}f_{j} B_{F}^{j} \left(x_{i} \right)  .                                              
\end{equation} 
For constraint $C_{3} $, let $\tau _{j} \left( \mathbf{X} \right)$ be a differentiation of storage characteristic of entangled states from the $j^{th} $ fidelity type: 
\begin{equation} \label{ZEqnNum854771} 
\tau _{j} \left( \mathbf{X} \right)=\sum _{i=1}^{N}\left(w_{j} \left(x_{i} \right)-\Omega \right)^{2}  , 
\end{equation} 
where 
\begin{equation} \label{18)} 
\Omega =\frac{\sum _{i=1}^{N}w_{j} \left(x_{i} \right) }{N} .                                                     
\end{equation} 
Then, $C_{3} $ is defined as
\begin{equation} \label{ZEqnNum901641} 
C_{3} :\nu \left( \mathbf{X} \right)\le \Pi ,                                                       
\end{equation} 
where $\Pi $ is an upper bound on the storage characteristic of entangled states from the $j^{th} $ fidelity type, while $\nu $ is evaluated via \eqref{ZEqnNum854771} as
\begin{equation} \label{20)} 
\nu =\sum _{j=1}^{N}\tau _{j} \left( \mathbf{X} \right).                                                        
\end{equation} 

\section{System Model}
\label{sec3}
 This section defines the Poisson entanglement optimization method, and it is applied to the solution of the multiobjective optimization problem of \sref{sec2}. 

\subsection{Motivation and Utility of the Mathematical Model in the Quantum Internet}
The quantum Internet is defined as a complex network model with quantum and classical layers that involve several optimization criteria and objectives. An optimization problem model of the quantum Internet therefore induces a multiobjective optimization problem model that considers the special requirements of the environment of the quantum Internet. These requirements cover the entanglement transmission procedure, processing of quantum entanglement in the quantum nodes, and auxiliary communication through the classical links that support the entangled network structure. The quantum transmission procedure models the generation of the entangled quantum network with quantitative and qualitative measures. In this manner, a quantitative measure is the relative entropy of entanglement between the quantum nodes, while the entanglement fidelity is a qualitative measure. Classical communication could also cause an overhead in the entanglement distribution mechanism of the quantum Internet. Thus, a multiobjective optimization framework should consider the attributes of both quantum and classical layers. 

To address the multiple criteria and several objectives of the quantum Internet, a multiobjective optimization framework is defined. The multiple criteria of the quantum Internet are defined as diverse objective functions that should be satisfied in parallel. The problem is therefore analogous to finding solutions in an objective space such that the objective space is defined and spanned by the input problems induced by the environment of the quantum Internet. The multiobjective optimization framework should evolve a set of solutions to the Pareto optimal front. In our model, these solutions are evolved via the mathematical model of epicenters that provide a naturally inspired answer to the multiobjective problem defined via the environment of the quantum Internet. The mathematical model of epicenters utilizes the theory of Pareto dominance in the problem resolution such that the selection and evaluation processes in the objective space that are required to identify a global optimal solution are controlled via our nature-inspired model. The proposed Poisson model ensures a robust randomization and efficient convergence in the objective space such that the solutions determined by utilizing the epicenters in the objective space will converge to a global optimal solution. The global optimal solution in the objective space represents the parallel satisfaction of the multiple criteria and objective functions defined by the quantum Internet. The randomness injected by the Poisson distribution not just avoids early convergence to a local optimal solution but also induces a fast convergence for the global optimal solution in the objective space. 

Since the multiple objectives and optimization criteria of the mathematical framework are motivated by practical assumptions and considerations of the quantum Internet, the proposed mathematical model of epicenters is strongly connected with a quantum Internet scenario. As follows, the utility of the proposed multiobjective optimization framework represents an effective solution for the practical optimization problems induced by the quantum Internet.

\subsection{Poisson Operators}

The attributes of the Poisson operator are as follows.
  
\subsubsection{Dispersion}
The $D\left({\rm {\mathcal E}}\right)$ dispersion coefficient of an epicenter ${\rm {\mathcal E}}$ (solution in the feasible space ${\rm {\mathcal S}}_{F} $) determines the number of affected $L_{j} $, $j=1,\ldots ,D\left({\rm {\mathcal E}}\right)$, locations around an epicenter ${\rm {\mathcal E}}$. The random locations around an epicenter also represent solutions in ${\rm {\mathcal S}}_{F} $ that help in increasing the diversity of population ${\rm {\mathcal P}}$ (a set of possible solutions) to find a global optimum. The diversity increment is therefore a tool to avoid an early convergence to a local optimum \cite{ref9, ref10}.

The dispersion $D\left({\rm {\mathcal E}}_{i} \right)$ operator for an $i^{th} $ epicenter ${\rm {\mathcal E}}_{i} $ is defined as
\begin{equation} \label{ZEqnNum754163} 
D\left( {{\mathcal{E}}_{i}} \right)=m\frac{\left( \tilde{f}\left( \langle \mathcal{E} \rangle   \right)-\tilde{f}\left( {{\mathcal{E}}_{i}} \right) \right)+\vartheta }{\sum\limits_{i=1}^{\left| \mathcal{P} \right|}{\left( \tilde{f}\left( \langle \mathcal{E} \rangle   \right)-\tilde{f}\left( {{\mathcal{E}}_{i}} \right) \right)+\vartheta }}, 
\end{equation} 
where $m$ is a control parameter, ${\rm {\mathcal E}}_{i} $ is an $i^{th} $ individual (epicenter) from the $\left|{\rm {\mathcal P}}\right|$ individuals (epicenters) in population ${\rm {\mathcal P}}$, $\left|{\rm {\mathcal P}}\right|$ is the size of population ${\rm {\mathcal P}}$, function $\tilde{f}\left(\cdot \right)$ is the fitness value (see \sref{app}), $\tilde{f}\left( \langle \mathcal{E} \rangle   \right)$ is a maximum objective value among the $\left|{\rm {\mathcal P}}\right|$ individuals, and $\vartheta $ is a residual quantity.

Without loss of generality, assuming $\left|{\rm {\mathcal P}}\right|$ epicenters, the $q$ total number of locations is as
\begin{equation} \label{ZEqnNum600378} 
q=\sum _{i=1}^{\left|{\rm {\mathcal P}}\right|}D\left({\rm {\mathcal E}}_{i} \right) .                                               
\end{equation} 
 
\subsubsection{Seismic Power and Magnitude}
Assume that $L_{j} $ is a random location around ${\rm {\mathcal E}}_{i} $. For $L_{j} $, the Euclidean distance $d\left({\rm {\mathcal E}}_{i} ,l_{j} \right)$ between an $i^{th} $ epicenter ${\rm {\mathcal E}}_{i} $ and the projection point $l_{j} $ of a $j^{th} $ location point $L_{j} $,$j=1,\ldots ,D\left({\rm {\mathcal E}}\right)$ on the ellipsoid around ${\rm {\mathcal E}}_{i} $ is as follows:
\begin{equation} \label{ZEqnNum881261} 
\begin{split} {d\left({\rm {\mathcal E}}_{i} ,l_{j} \right)} &= {\sqrt{\left(\dim _{1} \left(l_{j} \right)\right)^{2} +\left(\dim _{2} \left(l_{j} \right)\right)^{2} } } \\ 
 &=\sqrt{\frac{1+tg^{2} \alpha _{{\rm {\mathcal E}}_{i} } \left(l_{j} \right)}{a^{-2} +tg^{2} \alpha _{{\rm {\mathcal E}}_{i} } \left(l_{j} \right)} } , 
\end{split} 
\end{equation} 
where $\dim _{i} \left(\cdot \right)$ is the $i^{th} $ dimension of $l_{j} $, and  
\begin{equation} \label{24)} 
\frac{\left(\dim _{1} \left(l_{j} \right)\right)^{2} }{a^{2} } +\frac{\left(\dim _{2} \left(l_{j} \right)\right)^{2} }{b^{2} } =1, 
\end{equation} 
where coefficients $a$ and $b$ define the shape of the ellipse around epicenter ${\rm {\mathcal E}}_{i} $ (see \fref{fig2}), while $\alpha _{{\rm {\mathcal E}}_{i} } \left(l_{j} \right)$ is an angle: 
\begin{equation} \label{25)} 
tg\alpha _{{\rm {\mathcal E}}_{i} } \left(l_{j} \right)=\frac{\dim _{2} \left(l_{j} \right)}{\dim _{1} \left(l_{j} \right)} .                                            
\end{equation} 
The seismic power $P\left({\rm {\mathcal E}}_{i} ,L_{j} \right)$ operator for an $i^{th} $ epicenter ${\rm {\mathcal E}}_{i} $ in a $j^{th} $ location point $L_{j} $,$j=1,\ldots ,D\left({\rm {\mathcal E}}_{i} \right)$ is defined as
\begin{equation} \label{ZEqnNum485148} 
P\left({\rm {\mathcal E}}_{i} ,L_{j} \right)=\left(\frac{1}{d\left({\rm {\mathcal E}}_{i} ,l_{j} \right)} M\left({\rm {\mathcal E}}_{i} ,L_{j} \right)\right)^{b_{1} } b_{0} e^{\sigma _{\ln P\left({\rm {\mathcal E}}_{i} ,L_{j} \right)} } ,                               
\end{equation} 
where $b_{0} $ and $b_{1} $ are regression coefficients, $\sigma _{\ln P\left({\rm {\mathcal E}}_{j} \right)} $ is the standard deviation \cite{ref14}, $M\left({\rm {\mathcal E}}_{i} ,L_{j} \right)$ is the seismic magnitude in a location $L_{j} $, and $l_{j} $ is the projection of $L_{j} $ onto the ellipsoid around ${\rm {\mathcal E}}_{i} $ \cite{ref14}.

Thus, at a given $L_{j} $ with $d\left({\rm {\mathcal E}}_{i} ,l_{j} \right)$ (\eqref{ZEqnNum881261}), from $P\left({\rm {\mathcal E}}_{i} ,L_{j} \right)$ (see \eqref{ZEqnNum485148}), the magnitude $M\left({\rm {\mathcal E}}_{i} ,L_{j} \right)$ between epicenter ${\rm {\mathcal E}}_{i} $ and location $L_{j} $ is evaluated as
\begin{equation} \label{ZEqnNum704162} 
M\left({\rm {\mathcal E}}_{i} ,L_{j} \right)=\left(P\left({\rm {\mathcal E}}_{i} ,L_{j} \right)\frac{1}{b_{0} e^{\sigma _{\ln P\left({\rm {\mathcal E}}_{i} ,L_{j} \right)} } } \right)^{\frac{1}{b_{1} } } d\left({\rm {\mathcal E}}_{i} ,l_{j} \right).                                  
\end{equation} 

\subsubsection{Cumulative Magnitude}
Let $L_{j}^{{\rm {\mathcal E}}_{i} } $ be the location point where the seismic power $P\left({\rm {\mathcal E}}_{i} ,L_{j}^{{\rm {\mathcal E}}_{i} } \right)$ is maximal for a given epicenter ${\rm {\mathcal E}}_{i} $. Let $P^{{\rm *}} \left({\rm {\mathcal E}}_{i} \right)$ be the maximal seismic power, 
\begin{equation} \label{ZEqnNum397433} 
P^{{\rm *}} \left({\rm {\mathcal E}}_{i} \right)=\mathop{\max }\limits_{\forall j} P\left({\rm {\mathcal E}}_{i} ,L_{j}^{{\rm {\mathcal E}}_{i} } \right).                                          
\end{equation} 
Assuming that $\left|{\rm {\mathcal P}}\right|$ epicenters, ${\rm {\mathcal E}}_{1,\ldots ,\left|{\rm {\mathcal P}}\right|} $ exist in the system, let identify by $P_{\max } \left({\rm {\mathcal E}}'\right)$ the epicenter ${\rm {\mathcal E}}'$ with a maximal seismic power among as
\begin{equation} \label{ZEqnNum751288} 
P_{\max } \left({\rm {\mathcal E}}'\right)=\mathop{\max }\limits_{\forall i} \left(P^{{\rm *}} \left({\rm {\mathcal E}}_{i} \right)\right),                                           
\end{equation} 
with magnitude $M\left({\rm {\mathcal E}}',L_{j}^{{\rm {\mathcal E}}'} \right)$, where $L_{j}^{{\rm {\mathcal E}}'} $ is the location point where the seismic power $P_{\max } \left({\rm {\mathcal E}}'\right)$ is maximal yielded for ${\rm {\mathcal E}}'$.

Then the $C\left({\rm {\mathcal E}}_{i} \right)$ cumulative magnitude for an epicenter ${\rm {\mathcal E}}_{i} $ is defined as
\begin{equation} \label{ZEqnNum235110} 
C\left({\rm {\mathcal E}}_{i} \right)={\rm {\mathcal M}}\frac{\left(\tilde{f}\left({\rm {\mathcal E}}_{i} \right)-\tilde{f}\left({\rm {\mathcal E}}'\right)\right)+\vartheta }{\sum _{i=1}^{\left|{\rm {\mathcal P}}\right|}\left(\tilde{f}\left({\rm {\mathcal E}}_{i} \right)-\tilde{f}\left({\rm {\mathcal E}}'\right)\right)+\vartheta  } , 
\end{equation} 
where ${\rm {\mathcal E}}'$ is the highest seismic power epicenter with magnitude $M\left({\rm {\mathcal E}}',L_{j}^{{\rm {\mathcal E}}'} \right)$, $\tilde{f}\left({\rm {\mathcal E}}'\right)$ is the minimum objective value among the $\left|{\rm {\mathcal P}}\right|$ epicenters, and ${\rm {\mathcal M}}$ is a control parameter defined as
\begin{equation} \label{ZEqnNum597526} 
{\rm {\mathcal M}}=\sum _{i=1}^{\left|{\rm {\mathcal P}}\right|}M\left({\rm {\mathcal E}}_{i} ,L_{j}^{{\rm {\mathcal E}}_{i} } \right) ,                                               
\end{equation} 
where $L_{j}^{{\rm {\mathcal E}}_{i} } $ provides the maximal seismic power for an $i^{th} $ epicenter ${\rm {\mathcal E}}_{i} $, functions $\tilde{f}\left({\rm {\mathcal E}}_{i} \right)$ and $\tilde{f}\left({\rm {\mathcal E}}'\right)$ are the fitness values (see \sref{app}) for the current epicenter ${\rm {\mathcal E}}_{i} $ and for the highest seismic power epicenter ${\rm {\mathcal E}}'$, and $\vartheta $ is a residual quantity.

\subsection{Distribution of Epicenters}
Assume that ${\rm {\mathcal E}}_{i} $ is a current epicenter (solution) and ${\rm {\mathcal R}}_{k} $ and ${\rm {\mathcal R}}_{l} $ are two random reference points around ${\rm {\mathcal E}}_{i} $. Using the $C\left({\rm {\mathcal E}}_{i} \right)$ cumulative seismic magnitude (see \eqref{ZEqnNum235110}) of an epicenter ${\rm {\mathcal E}}_{i} $, the generation of a new epicenter ${\rm {\mathcal E}}_{p} $ is as follows:

Let $\Phi \left({\rm {\mathcal E}}_{i} ,{\rm {\mathcal R}}_{k} ,{\rm {\mathcal R}}_{l} \right)$ be a Poisson range identifier function \cite{ref12,ref13} for ${\rm {\mathcal E}}_{i} $ using ${\rm {\mathcal R}}_{k} $ and ${\rm {\mathcal R}}_{l} $ as random reference points:
\begin{equation} \label{ZEqnNum748170} 
\begin{split}
&\Phi \left({\rm {\mathcal E}}_{i} ,{\rm {\mathcal R}}_{k} ,{\rm {\mathcal R}}_{l} \right)\\&=\frac{d\left({\rm {\mathcal E}}_{i} ,{\rm {\mathcal R}}_{k} \right)c_{w} \left({\rm {\mathcal R}}_{k} ,{\rm {\mathcal R}}_{l} \right)}{\cos \left(\theta \left(\ell _{{\rm {\mathcal E}}_{i} ,{\rm {\mathcal R}}_{k} } ,\ell _{{\rm {\mathcal R}}_{k} ,{\rm {\mathcal R}}_{l} } \right)\right)\cdot d\left({\rm {\mathcal R}}_{k} ,{\rm {\mathcal R}}_{l} \right)c_{w} \left({\rm {\mathcal E}}_{i} ,{\rm {\mathcal R}}_{k} \right)} , 
\end{split}
\end{equation} 
where ${\rm {\mathcal E}}_{i} $ is a current epicenter, ${\rm {\mathcal R}}_{k} $ and ${\rm {\mathcal R}}_{l} $ are random reference points, $d\left(\cdot \right)$ is the Euclidean distance function, $c_{w} \left({\rm {\mathcal E}}_{i} ,{\rm {\mathcal R}}_{k} \right)$ and $c_{w} \left({\rm {\mathcal R}}_{k} ,{\rm {\mathcal R}}_{l} \right)$ are weighting coefficients between epicenters ${\rm {\mathcal E}}_{i} $ and ${\rm {\mathcal R}}_{k} $ and between ${\rm {\mathcal R}}_{k} $ and ${\rm {\mathcal R}}_{l} $, and $\theta \left(\ell _{{\rm {\mathcal E}}_{i} ,{\rm {\mathcal R}}_{k} } ,\ell _{{\rm {\mathcal R}}_{k} ,{\rm {\mathcal R}}_{l} } \right)$ is the angle between lines $\ell _{{\rm {\mathcal E}}_{i} ,{\rm {\mathcal R}}_{k} } $ and $\ell _{{\rm {\mathcal R}}_{k} ,{\rm {\mathcal R}}_{l} } $:
\begin{equation} \label{33)} 
\begin{split}
&\theta \left(\ell _{{\rm {\mathcal E}}_{i} ,{\rm {\mathcal R}}_{k} } ,\ell _{{\rm {\mathcal R}}_{k} ,{\rm {\mathcal R}}_{l} } \right)\\&=\cos ^{-1} \left(\frac{d\left({\rm {\mathcal E}}_{i} ,{\rm {\mathcal R}}_{k} \right)^{2} +d\left({\rm {\mathcal E}}_{k} ,{\rm {\mathcal R}}_{l} \right)^{2} -d\left({\rm {\mathcal E}}_{i} ,{\rm {\mathcal R}}_{l} \right)^{2} }{2d\left({\rm {\mathcal E}}_{i} ,{\rm {\mathcal R}}_{k} \right)d\left({\rm {\mathcal R}}_{k} ,{\rm {\mathcal R}}_{l} \right)} \right).                      
\end{split}
\end{equation} 
Without loss of generality, using \eqref{ZEqnNum748170}, a Poissonian distance function $\mathfrak{D}\left( {{\mathcal{E}}_{p}} \right)$ for the finding of new epicenter ${\rm {\mathcal E}}_{p} $ is defined via a $P$ Poisson distribution \cite{ref12,ref13} as follows:
\begin{equation} \label{34)} 
\mathfrak{D}\left( {{\mathcal{E}}_{p}} \right)=P\left(k,\lambda \right), 
\end{equation} 
where 
\begin{equation} \label{35)} 
k=\Phi \left({\rm {\mathcal E}}_{i} ,{\rm {\mathcal R}}_{k} ,{\rm {\mathcal R}}_{l} \right),                                                
\end{equation} 
with mean
\begin{equation} \label{36)} 
\lambda =\mathbb{E}\left[ \Phi \left( {{\mathcal{E}}_{i}},{{\mathcal{R}}_{k}},{{\mathcal{R}}_{l}} \right) \right].                                            
\end{equation} 
Therefore, the resulting new epicenter ${\rm {\mathcal E}}_{p} $ is a Poisson random epicenter ${\rm {\mathcal E}}_{p} $ with a Poisson range identifier $\mathfrak{D}\left( {{\mathcal{E}}_{p}} \right)$. 

For a large set of reference points, only those reference points that are within the $r\left({\rm {\mathcal E}}_{i} \right)$ radius around the current solution ${\rm {\mathcal E}}_{i} $ are selected for the determination of the new solution ${\rm {\mathcal E}}_{p} $. This radius is defined as
\begin{equation} \label{ZEqnNum593818} 
r\left({\rm {\mathcal E}}_{i} \right)=\chi 10^{Q_{1} \left(2\tilde{{\rm {\mathcal M}}}\right)-Q_{2} } ,                                              
\end{equation} 
where $\tilde{{M}}$ is the average magnitude, 
\begin{equation} \label{38)} 
\tilde{{M}}=\frac{1}{\left|{\rm {\mathcal P}}\right|} {\rm {\mathcal M}}=\frac{1}{\left|{\rm {\mathcal P}}\right|} \sum _{i=1}^{\left|{\rm {\mathcal P}}\right|}M\left({\rm {\mathcal E}}_{i} ,L_{j}^{{\rm {\mathcal E}}_{i} } \right) ,                                       
\end{equation} 
$Q_{1} $ and $Q_{2} $ are constants, and $\chi $ is a normalization term. Motivated by the corresponding seismologic relations of the Dobrovolsky-Megathrust radius formula \cite{ref13}, the constants in \eqref{ZEqnNum593818} are selected as $Q_{1} =0.414$ and $Q_{2} =1.696$. 

In the relevance range $r\left({\rm {\mathcal E}}_{i} \right)$ of \eqref{ZEqnNum593818}, the weights of reference points are determined by the seismic power function \eqref{ZEqnNum485148}.
 
\subsection{Population Diversity}
\subsubsection{Hypocentral}
The hypocentral of an epicenter is aimed to increase the diversity of population by a randomization. 

Let $\dim _{k} \left({\rm {\mathcal E}}_{i} \right)$ be an actual randomly selected $k^{th} $ dimension and $k=1,\ldots ,\dim \left({\rm {\mathcal E}}_{i} \right)$ be a current epicenter ${\rm {\mathcal E}}_{i} $, $i=1,\ldots ,\left|{\rm {\mathcal P}}\right|$. The ${\rm {\mathcal H}}\left(\dim _{k} \left({\rm {\mathcal E}}_{i} \right)\right)$ hypocentral provides a random displacement \cite{ref12,ref13} of $\dim _{k} \left({\rm {\mathcal E}}_{i} \right)$ using $C\left({\rm {\mathcal E}}_{i} \right)$ (see \eqref{ZEqnNum235110}):
\begin{equation} \label{ZEqnNum840814} 
\begin{split}
{\rm {\mathcal H}}\left(\dim _{k} \left({\rm {\mathcal E}}_{i} \right)\right)&=\dim _{k} \left({\rm {\mathcal E}}_{i'} \right) \\&=\sqrt{\begin{array}{l} {\left(\frac{1}{M\left(\dim _{k} \left({\rm {\mathcal E}}_{i} \right),L_{j}^{\dim _{k} \left({\rm {\mathcal E}}_{i} \right)} \right)} \dim _{k} \left({\rm {\mathcal E}}_{i} \right)\right)^{2} } \\ {+\left({\rm {\mathcal U}}\left(-C\left({\rm {\mathcal E}}_{i} \right),C\left({\rm {\mathcal E}}_{i} \right)\right)\right)^{2} } \end{array}}  
\end{split}
\end{equation} 
where ${\rm {\mathcal U}}\left(-C\left({\rm {\mathcal E}}_{i} \right),C\left({\rm {\mathcal E}}_{i} \right)\right)$ is a uniform random number from the range of $\left[-C\left({\rm {\mathcal E}}_{i} \right),C\left({\rm {\mathcal E}}_{i} \right)\right]$ to yield the displacement $\dim _{k} \left({\rm {\mathcal E}}_{i'} \right)$, $M\left(\dim _{k} \left({\rm {\mathcal E}}_{i} \right),L_{j}^{\dim _{k} \left({\rm {\mathcal E}}_{i} \right)} \right)$ is the magnitude, and $L_{j}^{\dim _{k} \left({\rm {\mathcal E}}_{i} \right)} $ is a location point where $P\left(\dim _{k} \left({\rm {\mathcal E}}_{i} \right),L_{j}^{\dim _{k} \left({\rm {\mathcal E}}_{i} \right)} \right)$ is maximal for $\dim _{k} \left({\rm {\mathcal E}}_{i} \right)$.

The $D\left({\rm {\mathcal E}}_{i} \right)$ locations around the cumulative magnitude $C\left({\rm {\mathcal E}}_{i} \right)$ of ${\rm {\mathcal E}}_{i} $ are generated by \eqref{ZEqnNum840814} through all the randomly selected $Y$ dimensions, where $Y$ is as follows \cite{ref9, ref10}:
\begin{equation} \label{40)} 
Y={\rm {\mathcal U}}\left(1,\dim \left({\rm {\mathcal E}}_{i} \right)\right).                                               
\end{equation} 
The process is repeated for all ${\rm {\mathcal E}}_{i} $.  
 
\subsubsection{Poisson Randomization}
To generate random locations around $\dim _{k} \left({\rm {\mathcal E}}_{i} \right)$, a Poisson distribution is also used to increase the diversity of the population. A random location in the $k^{th} $ dimension $L_{r}^{\dim _{k} \left({\rm {\mathcal E}}_{i} \right)} $ around $\dim _{k} \left({\rm {\mathcal E}}_{i} \right)$ is generated as follows:
\begin{equation} \label{ZEqnNum379228} 
L_{r}^{\dim _{k} \left({\rm {\mathcal E}}_{i} \right)} =\dim _{k} \left({\rm {\mathcal E}}_{i} \right)w,                                                
\end{equation} 
where 
\begin{equation} \label{42)} 
w\in P\left(X=k,\lambda \right) 
\end{equation} 
is a Poisson random number with distribution coefficients $k$ and $\lambda $. Given that it is possible that using \eqref{ZEqnNum379228} some randomly generated locations will be out of the feasible space ${\rm {\mathcal S}}_{F} $, a normalization operator ${\rm N} \left(\cdot \right)$ of $L_{r}^{\dim _{k} \left({\rm {\mathcal E}}_{i} \right)} $ is defined to keep the new locations around $\dim _{k} \left({\rm {\mathcal E}}_{i} \right)$ in ${\rm {\mathcal S}}_{F} $, as follows \cite{ref9,ref10}:
\begin{equation} \label{43)} 
L_{r}^{{{\dim}_{k}}\left( {{\mathcal{E}}_{i}} \right)}=L_{r}^{{{\dim}_{k}}\left( {{\mathcal{E}}_{i}} \right)}\left( \bmod \left( B_{up}^{k}-B_{low}^{k} \right) \right)+B_{low}^{k},
\end{equation} 
where $B_{low}^{k} $ and $B_{up}^{k} $ are lower and upper bounds on the boundaries of locations in a $k^{th} $ dimension, and $\bmod (\cdot )$ is to a modular arithmetic function. The procedure is repeated for the randomly selected $t={\rm {\mathcal U}}\left(1,\dim \left({\rm {\mathcal E}}_{i} \right)\right)$ dimensions of ${\rm {\mathcal E}}_{i} $, for $\forall i$. 

\subsection{Iterative Convergence}
The method of convergence of solutions in the Poisson optimization is summarized in Method 1.

\begin{method}
  \DontPrintSemicolon
\caption{Convergence of Solutions}
\textbf{Step 1.} Generate $\left|{\rm {\mathcal P}}\right|$ epicenters, ${\rm {\mathcal E}}_{1} ,\ldots ,{\rm {\mathcal E}}_{\left|{\rm {\mathcal P}}\right|} $, with $D\left({\rm {\mathcal E}}_{i} \right)$ random locations around a given $i^{th} $ epicenter ${\rm {\mathcal E}}_{i} $. 

\textbf{Step 2.} Select an epicenter ${\rm {\mathcal E}}_{i} $, and determine the seismic operators $D\left({\rm {\mathcal E}}_{i} \right)$, $P\left({\rm {\mathcal E}}_{i} ,L_{j} \right)$,$M\left({\rm {\mathcal E}}_{i} ,L_{j} \right)$. 

\textbf{Step 3.} Determine the $\mathfrak{D}\left( {{\mathcal{E}}_{p}} \right)$ Poisson distance function using references ${\rm {\mathcal R}}_{k} $ and ${\rm {\mathcal R}}_{l} $ to yield a new solution ${\rm {\mathcal E}}_{p} $.

\textbf{Step 4.} Repeat steps 1--3, until the closest epicenter to the ${\rm {\mathcal E}}'$ optimal epicenter is not found or other stopping criteria are not met. 
\end{method}
 
\begin{center}
\begin{figure*}[!h]
\begin{center}
\includegraphics[angle = 0,width=1 \linewidth]{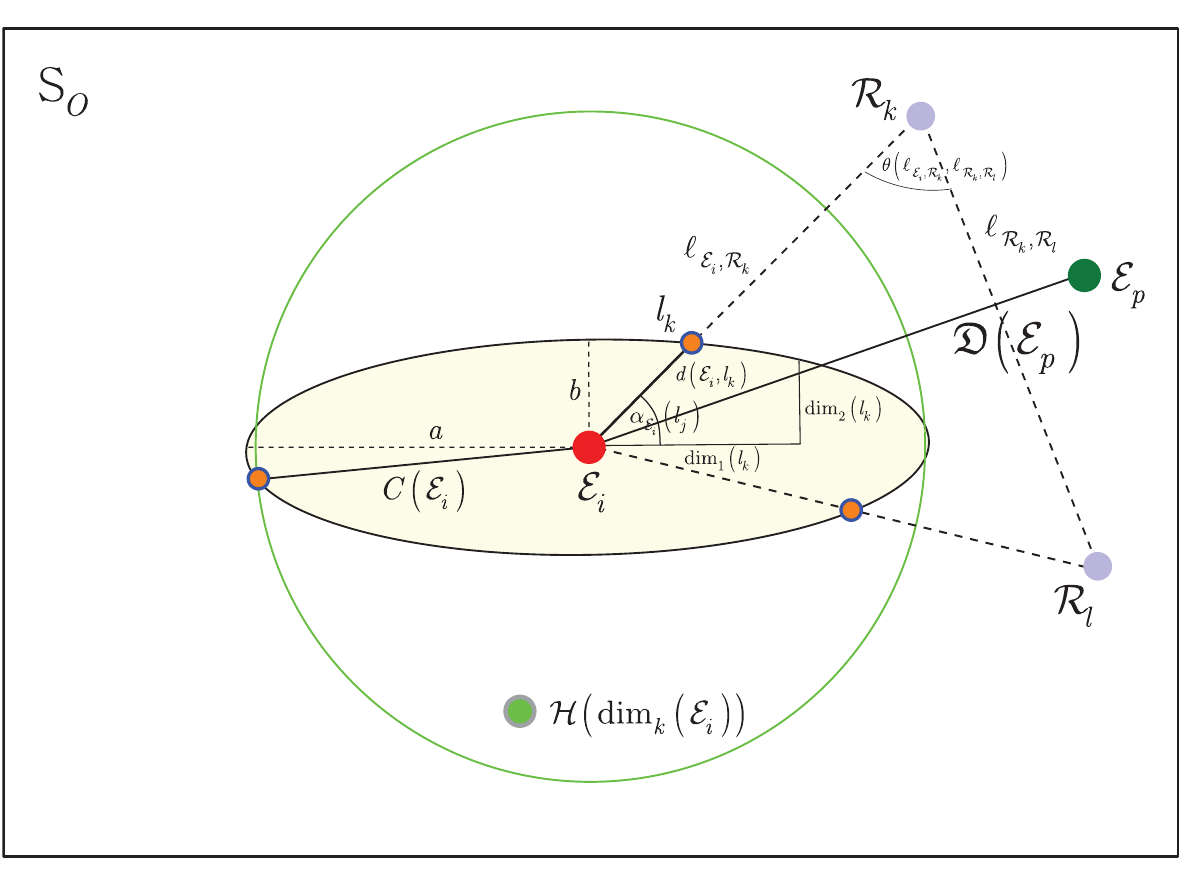}
\caption{Iteration step of the Poisson optimization model in the objective space ${\rm S}_{O} $. An $i^{th} $ epicenter, ${\rm {\mathcal E}}_{i} $ (depicted by the red dot), with a projected point $l_{k} $  of random reference location ${\rm {\mathcal R}}_{k} $. Reference locations ${\rm {\mathcal R}}_{k} $ and ${\rm {\mathcal R}}_{l} $ (blue dots) identify locations $L_{k} $ and $L_{l} $, respectively. The power in ${\rm {\mathcal R}}_{k} $ is $P\left({\rm {\mathcal E}}_{i} ,{\rm {\mathcal R}}_{k} \right)$ (see \eqref{ZEqnNum485148}), while the magnitude is $M\left({\rm {\mathcal E}}_{i} ,{\rm {\mathcal R}}_{k} \right)$ (see \eqref{ZEqnNum704162}). Notation $\dim _{i} \left(\cdot \right)$ refers to the $i^{th} $ dimension of $l_{k} $, and coefficients $a$ and $b$ define the shape of the ellipse (yellow) around epicenter ${\rm {\mathcal E}}_{i} $. The ${\rm {\mathcal H}}\left(\dim _{k} \left({\rm {\mathcal E}}_{i} \right)\right)$ hypocentral of ${\rm {\mathcal E}}_{i} $ is determined via the range of the $C\left({\rm {\mathcal E}}_{i} \right)$ cumulative magnitude (depicted by the green circle). The new epicenter ${\rm {\mathcal E}}_{p} $ (depicted by the green dot) is determined by the $\mathfrak{D}\left( {{\mathcal{E}}_{p}} \right)$ Poisson distance function using ${\rm {\mathcal R}}_{k} $ and ${\rm {\mathcal R}}_{l} $, with angle $\theta \left(\ell _{{\rm {\mathcal E}}_{i} ,{\rm {\mathcal R}}_{k} } ,\ell _{{\rm {\mathcal R}}_{k} ,{\rm {\mathcal R}}_{l} } \right)$ between lines $\ell _{{\rm {\mathcal E}}_{i} ,{\rm {\mathcal R}}_{k} } $ and $\ell _{{\rm {\mathcal R}}_{k} ,{\rm {\mathcal R}}_{l} } $.} 
 \label{fig2}
 \end{center}
\end{figure*}
\end{center}

An epicenter ${\rm {\mathcal E}}_{i} $ and the generation of a new solution ${\rm {\mathcal E}}_{p} $ with an in the objective space ${\rm S}_{O} $ are depicted in \fref{fig2}. The ellipsoid around ${\rm {\mathcal E}}_{i} $ and the projection point $l_{k} $ of the reference location ${\rm {\mathcal R}}_{k} $ are serving the determination of power function $P\left({\rm {\mathcal E}}_{i} ,{\rm {\mathcal R}}_{k} \right)$ in the reference location ${\rm {\mathcal R}}_{k} $.

A new epicenter ${\rm {\mathcal E}}_{p} $ is determined via the Poisson function $\mathfrak{D}\left( {{\mathcal{E}}_{p}} \right)$. Locations with low power function \eqref{ZEqnNum485148} values have high magnitudes \eqref{ZEqnNum704162} from the epicenter, whereas locations with high power function values have low magnitudes from the epicenter. 

\subsection{Framework}

The algorithmical framework that utilizes the Poisson entanglement optimization method for the problem statement presented in \sref{sec2} is defined in Algorithm 1.

\setcounter{algocf}{0}
\begin{algorithm}
\DontPrintSemicolon    
\caption{Poisson Optimization in the Quantum Internet}
\textbf{Step 0.} In an initial phase, a random population ${\rm {\mathcal P}}$ of $\left|{\rm {\mathcal P}}\right|$ feasible solutions is generated \cite{ref9, ref10} Let $G$ be an upper bound on the number of generations, $n_{G} $.

\textbf{Step 1.} For each epicenter $x_{i} ={\rm {\mathcal E}}_{i} $ in ${\rm {\mathcal P}}$, define $D\left({\rm {\mathcal E}}_{i} \right)$ random locations around ${\rm {\mathcal E}}_{i} $. For a diversity increment, determine the ${\rm {\mathcal H}}\left(\dim _{k} \left({\rm {\mathcal E}}_{i} \right)\right)$ hypocentral displacement function \eqref{ZEqnNum840814} for $\dim _{k} \left({\rm {\mathcal E}}_{i} \right)$, for $k=1,\ldots ,\dim \left({\rm {\mathcal E}}_{i} \right)$. 

 \textbf{Step 2.} Determine the seismic power $P\left({\rm {\mathcal E}}_{i} ,L_{j} \right)$ operator via \eqref{ZEqnNum485148} for an $i^{th} $ epicenter ${\rm {\mathcal E}}_{i} $ in a $j^{th} $ location point $L_{j} $,$j=1,\ldots ,D\left({\rm {\mathcal E}}_{i} \right)$. Determine the $L_{j}^{{\rm {\mathcal E}}_{i} } $, the location point where the seismic power $P\left({\rm {\mathcal E}}_{i} ,L_{j}^{{\rm {\mathcal E}}_{i} } \right)$ is maximal for a given epicenter ${\rm {\mathcal E}}_{i} $, via \eqref{ZEqnNum397433}. 

 \textbf{Step 3.} Determine epicenter ${\rm {\mathcal E}}'$ with a maximal seismic power $P_{\max } \left({\rm {\mathcal E}}'\right)$ via \eqref{ZEqnNum751288}. Compute seismic magnitude $M\left({\rm {\mathcal E}}',L_{j}^{{\rm {\mathcal E}}'} \right)$ via \eqref{ZEqnNum704162}, and determine the sum of all $N$ seismic magnitudes ${\rm {\mathcal M}}$ via \eqref{ZEqnNum597526}.

 \textbf{Step 4.} Compute the $D\left({\rm {\mathcal E}}_{i} \right)$ dispersion via \eqref{ZEqnNum754163} and the $C\left({\rm {\mathcal E}}_{i} \right)$ cumulative seismic magnitude via \eqref{ZEqnNum235110}. Select non-dominated solutions from the ${\rm {\mathcal P}}$ population set to the set ${\rm {\mathcal N}{\mathcal P}}$ of non-dominated solutions. Identify $\varphi _{k} $ as $\varphi _{k} =L_{k}^{{\rm {\mathcal E}}_{i} } $, where $L_{k}^{{\rm {\mathcal E}}_{k} } $ is a $k^{th} $ location around ${\rm {\mathcal E}}_{i} $. Update ${\rm {\mathcal N}{\mathcal P}}$ with the non-dominated solutions. 

 \textbf{Step 5.} Create set ${\rm {\mathcal P}}'$ of epicenters by selecting $p$ feasible solutions from ${\rm {\mathcal P}}$ using the $\Pr \left(\varphi _{i} \right)$ selection probability as $\Pr \left(\varphi _{i} \right)={\tilde{f}\left(\varphi _{i} \right) \mathord{\left/{\vphantom{\tilde{f}\left(\varphi _{i} \right) \sum _{r\in {\rm {\mathcal P}}}\tilde{f}\left(\varphi _{r} \right) }}\right.\kern-\nulldelimiterspace} \sum _{r\in {\rm {\mathcal P}}}\tilde{f}\left(\varphi _{r} \right) } $. Apply Sub-procedure 1. 

\textbf{Step 6.} If $n_{G} \ge G$, then stop the iteration; otherwise, repeat steps 1--4. 
\end{algorithm}

Sub-procedure 1 of step 5 is discussed in the Appendix. 
 
\subsubsection{Optimization of Classical Communications}
To achieve the minimization of classical communications required by the entanglement optimization, the $S$-metric (or hypervolume indicator) is integrated, which is a quality measure for the solutions or a contribution of a single solution in a solution set \cite{ref9, ref10} By definition, this metric identifies the size of dominated space (size of space covered).

By theory, the $S\left({\rm {\mathcal R}}\right)$ $S$-metric for a solution set ${\rm {\mathcal R}}=\left\{r_{1} ,\ldots ,r_{n} \right\}$ is as follows:
\begin{equation} \label{44)} 
S\left({\rm {\mathcal R}}\right)={\rm {\mathcal L}}\left(\bigcup _{r\in {\rm {\mathcal R}}}\left\{x_{ref} \angle x\angle \left. x\right|r\right\} \right),                                       
\end{equation} 
where ${\rm {\mathcal L}}$ is a Lebesgue measure, notation $b\angle a$ means $a$ dominates $b$ (or $b$ is dominated by $a$), and $x_{ref} $ is a reference point dominated by all valid solutions in the solution set \cite{ref9, ref10}. 

For a given solution $r_{i} $, the $S$-metric identifies the size of space dominated by $r_{i} $ but not dominated by any other solution, without loss of generality as:
\begin{equation} \label{45)} 
S\left( {{r}_{i}} \right)=\Delta S\left( \mathcal{R},{{r}_{i}} \right)=S\left( \mathcal{R} \right)-S\left( \mathcal{R}\text{ }\!\!\backslash\!\!\text{ }\left\{ {{r}_{i}} \right\} \right).                             
\end{equation} 
In the optimization of classical communications, the existence of two objective functions is assumed. The first objective function, $f_{1} $, is associated with the minimization of the cost of the first type of classical communications related to the reception and storage of entangled systems in the quantum nodes (it covers the classical communications related to the required entanglement throughput by the nodes, fidelity of received entanglement, number of stored entangled states, and fidelity parameters). Thus,
\begin{equation} \label{46)} 
f_{1} :\mathop{\min }\limits_{\forall i} {\rm {\mathcal C}}_{1} \left(x_{i} \right),                                                 
\end{equation} 
where ${\rm {\mathcal C}}_{1} \left(x_{i} \right)$ is the cost associated with the first type of classical communications related to a $x_{i} $.

The second objective function, $f_{2} $, is associated with the cost of the second type of classical communications that is related to entanglement purification:
\begin{equation} \label{47)} 
f_{2} :\mathop{\min }\limits_{\forall i} {\rm {\mathcal C}}_{2} \left(x_{i} \right),                                                  
\end{equation} 
where ${\rm {\mathcal C}}_{2} \left(x_{i} \right)$ is the cost associated with the second type of classical communications with respect to $x_{i} $.

Assuming objective functions $f_{1} $ and $f_{2} $, the $S\left(r_{i} \right)$ of a particular solution $r_{i} $ is as follows:
\begin{equation} \label{ZEqnNum180032} 
S\left(r_{i} \right)=\left(f_{1} \left(r_{i} \right)-f_{1} \left(r_{i-1} \right)\right)\left(f_{2} \left(r_{i} \right)-f_{2} \left(r_{i+1} \right)\right).                             
\end{equation} 
Given that the $S$-metric is calculated for the solutions, a set of nearest neighbors that restrict the space can be determined. Since the volume of this space can be quantified by the hypervolume, the solutions that satisfy objectives $f_{1} $ and $f_{2} $ can be found by utilizing \eqref{ZEqnNum180032}. 

\subsection{Computational Complexity}
The computational complexity of the Poissonian optimization method is derived as follows. Given that $\left|{\rm {\mathcal P}}\right|$ epicenters are generated in the search space and that the number of locations for an $i^{th} $ epicenter ${\rm {\mathcal E}}_{i} $ is determined by the dispersion operator $D\left({\rm {\mathcal E}}_{i} \right)$, the resulting computational complexity at a total number of locations $q=\sum _{i=1}^{\left|{\rm {\mathcal P}}\right|}D\left({\rm {\mathcal E}}_{i} \right) $ (see \eqref{ZEqnNum600378}) is 
\begin{equation} \label{49)} 
{\rm {\mathcal O}}\left(\left(\left|{\rm {\mathcal P}}\right|+q\right)^{{d \mathord{\left/{\vphantom{d 2}}\right.\kern-\nulldelimiterspace} 2} } \log \left(\left|{\rm {\mathcal P}}\right|+q\right)\right),                                            
\end{equation} 
since after a sorting process the locations for a given epicenter ${\rm {\mathcal E}}_{i} $ can be calculated with complexity ${\rm {\mathcal O}}\left(D\left({\rm {\mathcal E}}_{i} \right)\right)$, where $d$ is the number of objectives. 

Considering that in our setting $d=2$, the total complexity is 
\begin{equation} \label{50)} 
{\rm {\mathcal O}}\left(\left(\left|{\rm {\mathcal P}}\right|+q\right)\log \left(\left|{\rm {\mathcal P}}\right|+q\right)\right).                                         
\end{equation} 

\section{Problem Resolution}
\label{sec4}
The resolution of the problem shown in \sref{sec2} using the Poissonian entanglement optimization framework of \sref{sec3} is as follows.

Let $X_{{\rm {\mathcal S}}_{low} } $ be a set of nodes for which condition \eqref{ZEqnNum766550} holds for the fidelity of the received entangled states in the nodes, and let $X_{{\rm {\mathcal S}}_{high} } $ be a set of nodes for which condition \eqref{ZEqnNum229590} holds for the received fidelity entanglement.

Then let $\left|X_{{\rm {\mathcal S}}_{low} } \right|$ and $\left|X_{{\rm {\mathcal S}}_{high} } \right|$ be the cardinality of $X_{{\rm {\mathcal S}}_{low} } $ and $X_{{\rm {\mathcal S}}_{high} } $, respectively. 

Specifically, function \eqref{ZEqnNum737898} for the $X_{{\rm {\mathcal S}}_{low} } $-type nodes is rewritten as
\begin{equation} \label{ZEqnNum484298} 
{{\mathcal{G}}^{{{X}_{{{\mathcal{S}}_{low}}}}}}\left( \mathbf{X} \right)=\max \sum\limits_{i=1}^{\left| {{X}_{{{\mathcal{S}}_{low}}}} \right|}{\mathcal{F}_{i}^{{{X}_{{{\mathcal{S}}_{low}}}}}\left( \mathbf{X} \right)\mathbb{E}\left( D_{i}^{{{X}_{{{\mathcal{S}}_{low}}}}}\left( \mathbf{X} \right) \right)}, 
\end{equation} 
where ${\rm {\mathcal F}}_{i}^{X_{{\rm {\mathcal S}}_{low} } } \left(X\right)$ is the entanglement fidelity function for an $i^{th} $ $X_{{\rm {\mathcal S}}_{low} } $-type node $x_{i} $, $x_{i} \in X_{{\rm {\mathcal S}}_{low} } $, and $\mathbb{E}\left( D_{i}^{{{X}_{{{\mathcal{S}}_{low}}}}}\left( \mathbf{X} \right) \right)$ is the expected relative entropy of entanglement in an $i^{th} $ $X_{{\rm {\mathcal S}}_{low} } $-type $x_{i} $.

Similarly, for the $X_{{\rm {\mathcal S}}_{low} } $-type nodes, function \eqref{ZEqnNum737898} is as follows:
\begin{equation} \label{ZEqnNum158316} 
{{\mathcal{G}}^{{{X}_{{{\mathcal{S}}_{high}}}}}}\left( \mathbf{X} \right)=\max \sum\limits_{i=1}^{\left| {{X}_{{{\mathcal{S}}_{high}}}} \right|}{\mathcal{F}_{i}^{{{X}_{{{\mathcal{S}}_{high}}}}}\left( \mathbf{X} \right)\mathbb{E}\left( D_{i}^{{{X}_{{{\mathcal{S}}_{high}}}}}\left( \mathbf{X} \right) \right)}.                       
\end{equation} 
From \eqref{ZEqnNum484298} and \eqref{ZEqnNum158316}, a cumulative ${{\mathcal{G}}^{{{X}_{{{\mathcal{S}}_{high}}}}\otimes {{X}_{{{\mathcal{S}}_{high}}}}}}\left( \mathbf{X} \right)$ is defined as
\begin{equation} \label{53)} 
\begin{split}
   {{\mathcal{G}}^{{{X}_{{{\mathcal{S}}_{low}}}}\otimes {{X}_{{{\mathcal{S}}_{high}}}}}}\left( \mathbf{X} \right)=&\sum\limits_{i=1}^{\left| {{X}_{{{\mathcal{S}}_{high}}}} \right|}{{{A}_{i}}\mathcal{F}_{i}^{{{X}_{{{\mathcal{S}}_{high}}}}}\left( \mathbf{X} \right)\mathbb{E}\left( D_{i}^{{{X}_{{{\mathcal{S}}_{high}}}}}\left( \mathbf{X} \right) \right)} \\ 
 & +\sum\limits_{i=\left| {{X}_{{{\mathcal{S}}_{high}}}} \right|+1}^{\left| {{X}_{{{\mathcal{S}}_{low}}}} \right|+\left| {{X}_{{{\mathcal{S}}_{high}}}} \right|}{{{A}_{i}}\mathcal{F}_{i}^{{{X}_{{{\mathcal{S}}_{low}}}}}\left( \mathbf{X} \right)\mathbb{E}\left( D_{i}^{{{X}_{{{\mathcal{S}}_{low}}}}}\left( \mathbf{X} \right) \right){{F}_{1}}\left( \mathbf{X} \right),}  
\end{split}
\end{equation} 
 where $A_{i} $ refers to the number of received entangled systems in an $i^{th} $ node, while 
\begin{equation} \label{54)} 
{{F}_{1}}\left( \mathbf{X} \right)=\min \mathcal{C}\left( \mathbf{X} \right)=\sum\limits_{i=1}^{N}{\sum\limits_{i=1}^{T}{{{f}_{j}}B_{F}^{j}\left( {{x}_{i}} \right)}}.                               
\end{equation} 
The fidelity types of the available resource states in the nodes should be further divided into $T$ classes. The final function is then evaluated as
\begin{equation} \label{55)} 
\begin{split}
{{\mathcal{G}}^{{{X}_{{{\mathcal{S}}_{low}}}}\otimes {{X}_{{{\mathcal{S}}_{high}}}}}}\left( \mathbf{X} \right)&={{F}_{1}}\left( \mathbf{X} \right){{F}_{2}}\left( \mathbf{X} \right)\\&=\sum\limits_{i=1}^{\left| {{X}_{{{\mathcal{S}}_{low}}}} \right|+\left| {{X}_{{{\mathcal{S}}_{high}}}} \right|}{\sum\limits_{j=1}^{T}{{{f}_{j}}B_{F}^{j}\left( {{x}_{i}} \right)}{{F}_{2}}\left( \mathbf{X} \right),}
\end{split}
\end{equation} 
where 
\begin{equation} \label{56)} 
{{F}_{2}}\left( \mathbf{X} \right)=\min \mathcal{S}\left( \mathbf{X} \right)=\sum\limits_{i=1}^{\left| {{X}_{{{\mathcal{S}}_{low}}}} \right|+\left| {{X}_{{{\mathcal{S}}_{high}}}} \right|}{{{\mathcal{S}}_{i}}\left( \mathbf{X} \right)}.                               
\end{equation} 
Thus, 
\begin{equation} \label{57)} 
{{\mathcal{G}}^{{{X}_{{{\mathcal{S}}_{low}}}}\otimes {{X}_{{{\mathcal{S}}_{high}}}}}}\left( \mathbf{X} \right)=\sum\limits_{i=1}^{\left| {{X}_{{{\mathcal{S}}_{low}}}} \right|+\left| {{X}_{{{\mathcal{S}}_{high}}}} \right|}{{{\mathcal{S}}_{i}}\left( \mathbf{X} \right)},                                   
\end{equation} 
such that \cite{ref9, ref10}
\begin{equation} \label{58)} 
\begin{split}
   \sum\limits_{i=1}^{\left| {{X}_{{{\mathcal{S}}_{low}}}} \right|+\left| {{X}_{{{\mathcal{S}}_{high}}}} \right|}{{{\mathcal{F}}_{i}}\left( \mathbf{X} \right)}&\ge \gamma {{F}_{1}}\left( N \right)\\&=\gamma \sum\limits_{i=1}^{\left| {{X}_{{{\mathcal{S}}_{low}}}} \right|+\left| {{X}_{{{\mathcal{S}}_{high}}}} \right|}{\sum\limits_{j=1}^{T}{{{f}_{j}}B_{F}^{j}\left( {{x}_{i}} \right)}} \\ 
 & \le {{\nu }_{\mathbf{X}}}\left( {{\varphi }_{i}} \right)\Lambda \le \Pi B_{F}^{j}\left( {{x}_{i}} \right),  
\end{split}
\end{equation} 
where ${{\nu }_{\mathbf{X}}}\left( {{\varphi }_{i}} \right)=\sum\limits_{j=1}^{Z}{{{\tau }_{j}}\left( {{\varphi }_{i}} \right)} $, $\gamma $ is given by the constraint of \eqref{ZEqnNum992191}, while $\Pi $ is given by the constraint of \eqref{ZEqnNum901641}.  

\subsection{Convergence of Solutions}
Let ${{\mathcal{F}}_{i}}\left( \mathbf{X} \right)\in \left[0,1\right]$ be the objective function that refers to the resulting entanglement fidelity in a particular node $x_{i} $, after purification and quantum error correction with per-node cost functions $F_{1}^{i} \left( \mathbf{X} \right)$, and $F_{2}^{i} \left( \mathbf{X} \right)$, respectively.

Precisely, a current $i^{th} $ epicenter ${\rm {\mathcal E}}_{i} $ identifies a solution in the objective space ${\rm S}_{O}$,
\begin{equation}
{\rm S}_{O} :\left\{F_{1}^{i} \left(N\right),F_{2}^{i} \left(N\right),{{\mathcal{F}}_{i}}\left( \mathbf{X} \right)\right\}.
\end{equation}
The random locations around ${\rm {\mathcal E}}_{i} $ also represent possible solutions. Let ${\rm {\mathcal E}}^{{\rm *}} $ be an optimal solution in the ${\rm S}_{O} $ subject space, which maximizes ${{\mathcal{F}}_{i}}\left( \mathbf{X} \right)$ and minimizes $F_{1}^{i} \left( \mathbf{X} \right)$ and $F_{2}^{i} \left( \mathbf{X} \right)$. From ${\rm {\mathcal E}}_{i} $, the algorithm determines a new solution (epicenter) ${\rm {\mathcal E}}_{p} $ via the $\mathfrak{D}\left( {{\mathcal{E}}_{p}} \right)$ Poisson distance function, using the connection model between the locations around ${\rm {\mathcal E}}_{i} $. To improve the diversity, locations around ${\rm {\mathcal E}}_{p} $ are generated. The new epicenter ${\rm {\mathcal E}}_{p} $ converges to an optimal solution ${\rm {\mathcal E}}^{{\rm *}} $. The iterations are repeated until ${\rm {\mathcal E}}^{{\rm *}} $ is not found or until a stopping criterion is met.

\begin{center}
\begin{figure}[!h]
\begin{center}
\includegraphics[angle = 0,width=1\linewidth]{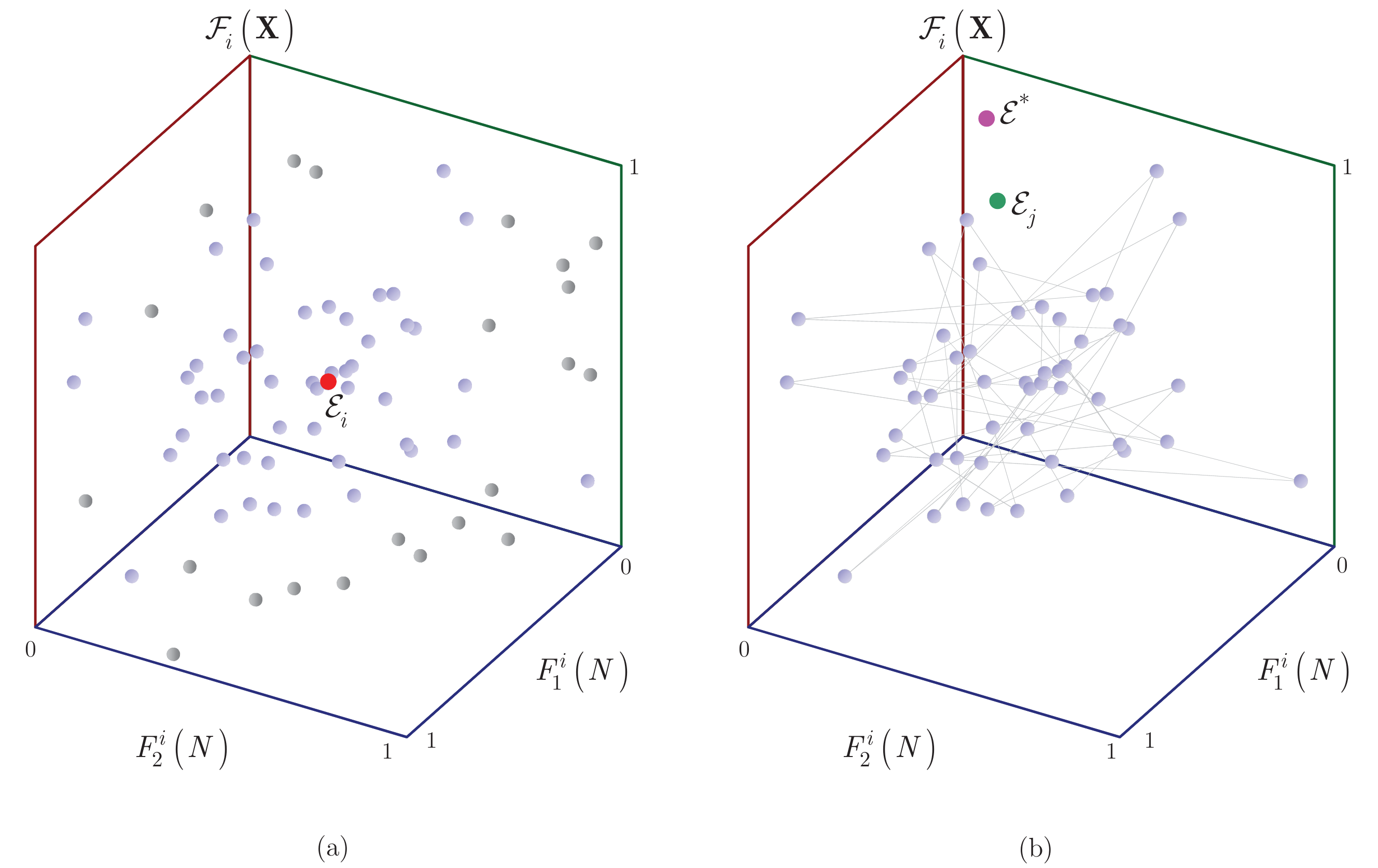}
\caption{Distribution of solutions for entanglement fidelity maximization in the objective space ${\rm S}_{O} :\left\{F_{1}^{i} \left(N\right),F_{2}^{i} \left(N\right),{{\mathcal{F}}_{i}}\left( \mathbf{X} \right)\right\}$ of the entanglement optimization problem (cost functions $F_{1} \left(N\right)$ and $F_{2} \left(N\right)$ and the objective function ${{\mathcal{F}}_{i}}\left( \mathbf{X} \right)$ are normalized onto the range of $\left[0,1\right]$). (a): A random epicenter ${\rm {\mathcal E}}_{i} $ refers to a current solution (depicted by the red dot) with the Poisson distributed reference locations (the reference points are not real solutions). The random reference locations are clustered into two classes: (1) reference locations within radius $r\left({\rm {\mathcal E}}_{i} \right)$ around ${\rm {\mathcal E}}_{i} $ and (2) reference locations outside the radius (depicted by the gray dots). Reference locations outside the range are neglected in the iteration. (b): A new epicenter ${\rm {\mathcal E}}_{p} $ (depicted by the green dot) is determined via the connection model of relevant reference points (e.g., lie inside the range of $r\left({\rm {\mathcal E}}_{i} \right)$), which yields the $\mathfrak{D}\left( {{\mathcal{E}}_{p}} \right)$ Poisson distance function. The new solution, ${\rm {\mathcal E}}_{p} $, converges toward an optimal solution ${\rm {\mathcal E}}^{{\rm *}} $ (depicted by the purple dot). The reference locations inside the relevance region are weighted by the seismic power function.} 
 \label{fig3}
 \end{center}
\end{figure}
\end{center}

The iteration from a current solution ${\rm {\mathcal E}}_{i} $ to a new solution ${\rm {\mathcal E}}_{p} $ toward a global optimal ${\rm {\mathcal E}}^{{\rm *}} $ in ${\rm S}_{O} $ is illustrated in \fref{fig3}. 

\section{Numerical Evidence}
\label{sec5}
In this section, a numerical evidence is proposed to demonstrate the Poisson entanglement optimization method. 
 
\subsection{Decision Making}

To demonstrate the results of \sref{sec4}, let ${{\mathcal{F}}_{i}}\left( \mathbf{X} \right)$ be the object function subject to maximize. The problem is to determine a matrix $\mathbf{X}$ that maximizes ${{\mathcal{F}}_{i}}\left( \mathbf{X} \right)$, and also $\mathbb{E}\left( D_{i}^{{{N}_{{{\mathcal{S}}_{low}}}}}\left( \mathbf{X} \right) \right)$, and minimizes the cost functions $F_{1}^{i} \left(N\right)$ and $F_{1}^{i} \left(N\right)$. Thus, for each node $N$, the optimal number of received and stored entangled systems should be determined, with high and low fidelity classes. 

Particularly, finding an optimal solution ${\rm {\mathcal E}}^{{\rm *}} $ in ${\rm S}_{O} $ with the assumptions given \sref{sec4}, is therefore means the selection of the optimal objective function (e.g., maximizing the entanglement fidelity ${{\mathcal{F}}_{i}}\left( \mathbf{X} \right)$ or maximizing the relative entropy of entanglement $\mathbb{E}\left( D_{i}^{{{N}_{{{\mathcal{S}}_{low}}}}}\left( \mathbf{X} \right) \right)$), in particular node types $X_{{\rm {\mathcal S}}_{low} } $ and $X_{{\rm {\mathcal S}}_{high} } $, while all cost functions are minimized in the quantum network. 

\begin{center}
\begin{figure}[!h]
\begin{center}
\includegraphics[angle = 0,width=1\linewidth]{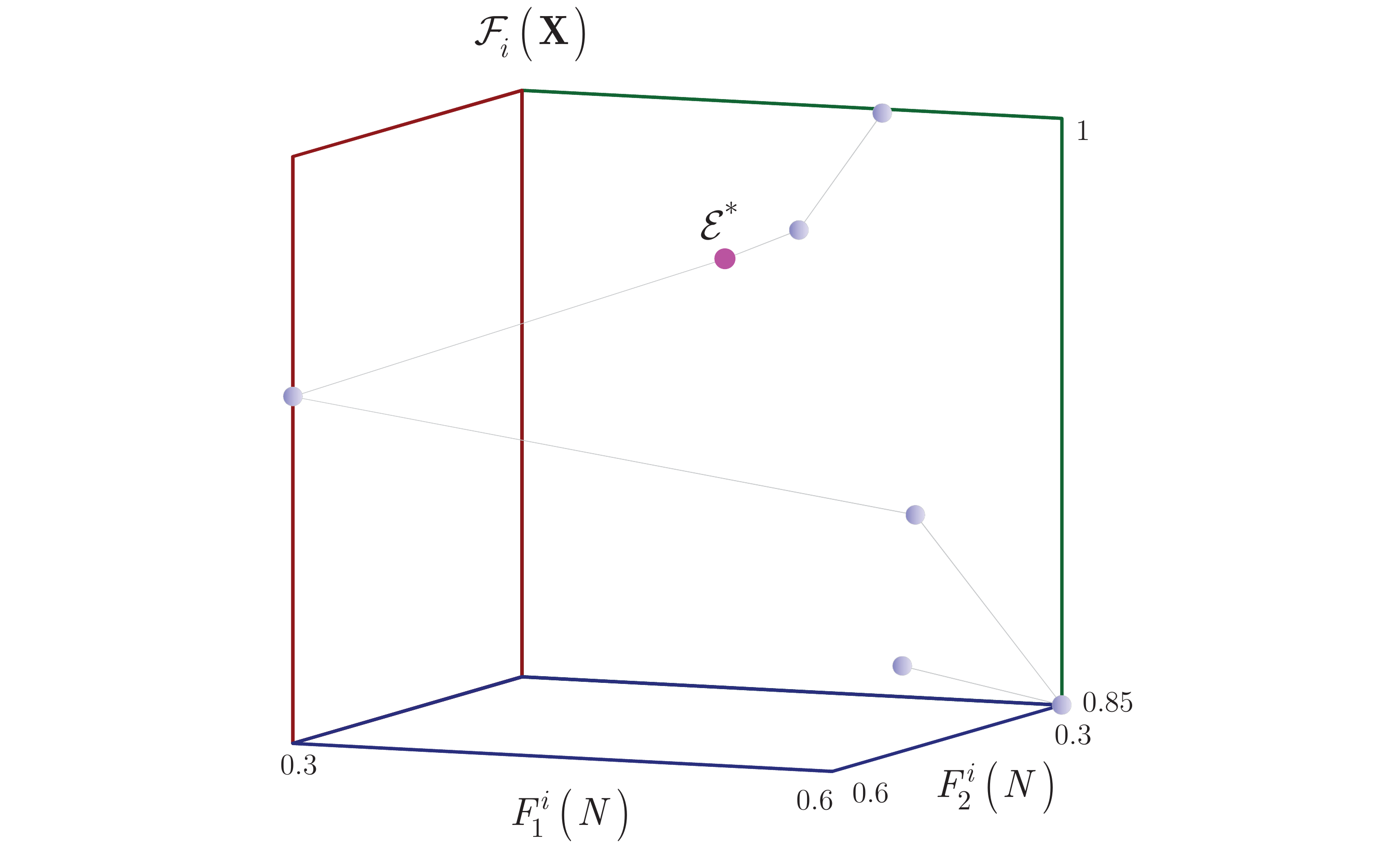}
\caption{Solution set in ${\rm S}_{O} $, with an optimal epicenter ${\rm {\mathcal E}}^{{\rm *}} $, $F_{1}^{i} \left(N\right)\in \left[0.3,0.6\right]$, $F_{2}^{i} \left(N\right)\in \left[0.3,0.6\right]$, ${{\mathcal{F}}_{i}}\left( \mathbf{X} \right)\in \left[0.85,1\right]$.} 
 \label{fig4}
 \end{center}
\end{figure}
\end{center}

A solution set in ${\rm S}_{O} $ is depicted in \fref{fig4}. 

An optimal solution ${\rm {\mathcal E}}^{{\rm *}} $ in ${\rm S}_{O} $ therefore yields the maximization of entanglement fidelity ${\rm {\mathcal F}}_{N} \left(X\right)$ if a particular node $N$ belongs to the class $N_{{\rm {\mathcal S}}_{high} } $, whereas it maximizes the relative entropy of entanglement $\mathbb{E}\left( D_{i}^{{{N}_{{{\mathcal{S}}_{low}}}}}\left( \mathbf{X} \right) \right)$ if $N$ belongs to the class $N_{{\rm {\mathcal S}}_{low} } $. Increasing $B_{F}^{j} \left(x_{i} \right)$ for a $N_{{\rm {\mathcal S}}_{high} } $-class node and then performing an optimal purification and quantum error correction could significantly improve the fidelity of entanglement. On the other hand, for a $N_{{\rm {\mathcal S}}_{low} } $-class node, the fidelity improvement at an optimal purification and quantum error correction is insignificant. Thus, incrementing $B_{F}^{j} \left(x_{i} \right)$ does not lead to a significant improvement in the fidelity. The optimal solution for these nodes is to focus on improving the relative entropy of entanglement, which requires lower cost function values.

This decision strategy provides a global optimal with respect to all quantum nodes of the quantum network. 

\begin{center}
\begin{figure}[!h]
\begin{center}
\includegraphics[angle = 0,width=1\linewidth]{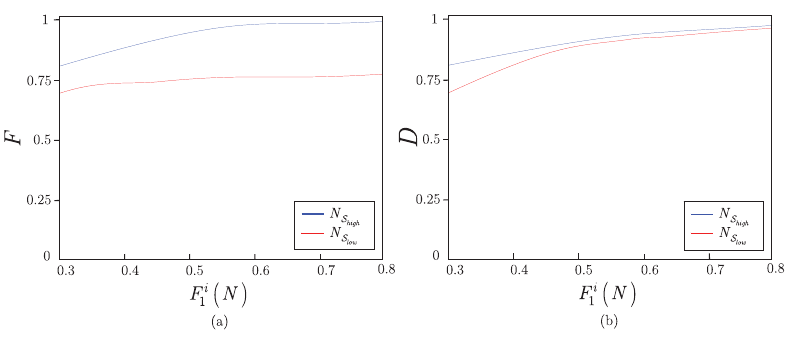}
\caption{Illustration of the decision making. (a): The $F$ entanglement fidelity values in function of $F_{1}^{i} \left(N\right)$ for $N_{{\rm {\mathcal S}}_{low} } $ (red line) and $N_{{\rm {\mathcal S}}_{high} } $ (blue line) nodes. The value of $F_{2}^{i} \left(N\right)$ is set to constant. (b): The $D$ relative entropy of entanglement values in function of $F_{1}^{i} \left(N\right)$ for $N_{{\rm {\mathcal S}}_{low} } $ (red line) and $N_{{\rm {\mathcal S}}_{high} } $(blue line) nodes. The value of $F_{2}^{i} \left(N\right)$ is set to constant.} 
 \label{fige1}
 \end{center}
\end{figure}
\end{center}

The decision making is illustrated in \fref{fige1}. In \fref{fige1}(a), the $F$  entanglement fidelity is depicted in function of $F_{1}^{i} \left(N\right)$ for $N_{{\rm {\mathcal S}}_{low} } $ and $N_{{\rm {\mathcal S}}_{high} } $ nodes. In \fref{fige1}(b), the $D$  relative entropy of entanglement is depicted in function of $F_{1}^{i} \left(N\right)$ for $N_{{\rm {\mathcal S}}_{low} } $ and $N_{{\rm {\mathcal S}}_{high} } $ nodes. The initial values of $F$ and $D$ are assumed to be equal for a given class, while the value of $F_{2}^{i} \left(N\right)$ is set to constant for illustration purposes.

For an $N_{{\rm {\mathcal S}}_{high} } $ node, the increment of $F_{1}^{i} \left(N\right)$ leads to significant improvement in $F$, while the increment in $D$ is moderate. For an $N_{{\rm {\mathcal S}}_{low} } $ node, the increment of $F_{1}^{i} \left(N\right)$ leads to moderate improvement in $F$, while the improvement in $D$ is significant. As a corollary, the increment of the entanglement throughput is a useful approach to increase the entanglement fidelity for the $X_{{\rm {\mathcal S}}_{high} } $ set, and to boost the relative entropy of entanglement in the $X_{{\rm {\mathcal S}}_{low} } $ set.

 \subsection{Distribution of Solutions}
First, we analyze the distribution of solutions in the feasible space ${\rm {\mathcal S}}_{F} $ focusing on the magnitudes associated to the locations around epicenters. 

Let assume that the total number of $q$ locations (see \eqref{ZEqnNum600378}) can be divided into $m$ magnitude ranges \cite{ref11}, such that
\begin{equation} \label{59)} 
\sum _{i=1}^{m}n_{i}  =q=\sum _{i=1}^{\left|{\rm {\mathcal P}}\right|}D\left({\rm {\mathcal E}}_{i} \right) ,                                             
\end{equation} 
where $n_{i} $ is the number of locations belonging to an $i^{th} $ magnitude range, $\left|{\rm {\mathcal P}}\right|$ is the population size. Then let $M_{i} $ be the magnitude associated to the $i^{th} $ magnitude range. Then a $\tilde{n}_{i} $ approximation of $n_{i} $ is evaluated as
\begin{equation} \label{60)} 
\tilde{n}_{i} =f\left(M_{i} \right),                                                    
\end{equation} 
where $f\left(\cdot \right)$ is a fitting function. To give an estimate on $n_{i} $ at a particular magnitude $M_{i} $, we utilize a power law distribution \cite{ref11} function ${\rm {\mathcal B}}\left(n_{i} \right)$ for a log-scaled $n_{i} $, as
\begin{equation} \label{ZEqnNum751356} 
{\rm {\mathcal B}}\left(n_{i} \right):\log _{10} \left(n_{i} \right)=a-b\tilde{M}_{i} ,                                        
\end{equation} 
where $\tilde{M}_{i} $ is a log scaled $M_{i} $, while $a$ and $b$ are constants \cite{ref11}. 

Then, the $\tilde{n}_{i} $ Poisson estimate is yielded as
\begin{equation} \label{62)} 
\tilde{n}_{i} =\sigma _{i}^{2} =\lambda _{i} ,                                                      
\end{equation} 
where $\sigma _{i}^{2} $ is the observational variance, while $\lambda _{i} $ is the mean of a Poisson distribution. Since the sum of independent Poisson variables is also a Poisson variable with mean equals to the sum of the components means, 
\begin{equation} \label{ZEqnNum747242} 
\lambda \left(q\right)=\sum _{i=1}^{m}\lambda _{i}  \approx q,                                                    
\end{equation} 
where $\lambda \left(q\right)$ is the mean total number, while $\lambda _{i} $ is an $i^{th} $ component mean. Using the Poisson property $\sigma ^{2} =\lambda $, the $\sigma _{q}^{2} $ estimated uncertainty is yielded as \cite{ref11} $\sigma _{q}^{2} =\lambda \left(q\right)=\sum _{i=1}^{m}f\left(\tilde{M}_{i} \right) $. Thus using a corresponding fitting function $f\left(\cdot \right)$, the mean and the variance of the total number of events are equal to the sum of the fitted values. 

In our model the distribution of the log scaled $\tilde{n}_{i} =\lambda _{i} $ values in function of $M_{i} $ are well approachable by the power law distribution ${\rm {\mathcal B}}\left(\lambda _{i} \right):\log _{10} \left(\lambda _{i} \right)=a-b\tilde{M}_{i} $, while the distribution of the $\lambda \left(q\right)$ total number \eqref{ZEqnNum747242} of locations are approachable by a ${\rm {\mathcal N}}\left(\lambda \left(q\right),\sigma _{{\rm {\mathcal N}}}^{2} \right)$, Gaussian distribution with variance $\sigma _{{\rm {\mathcal N}}}^{2} =\lambda \left(q\right)$ as $\lambda \left(q\right)\to \infty $, by theory.

\begin{center}
\begin{figure*}[!h]
\begin{center}
\includegraphics[angle = 0,width=0.8\linewidth]{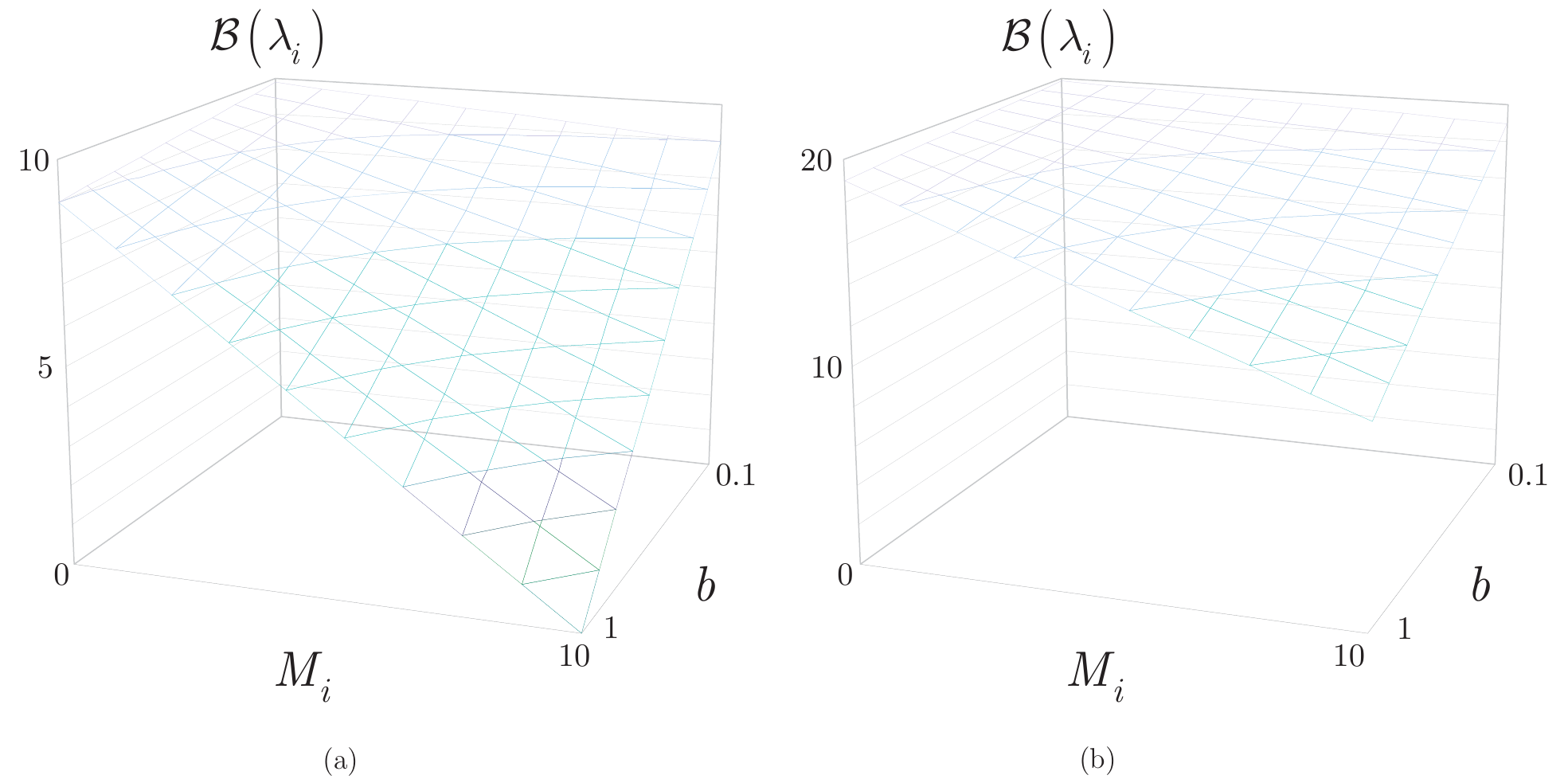}
\caption{Distribution of ${\rm {\mathcal B}}\left(\lambda _{i} \right)$ for different magnitudes $M_{i} $, $M_{i} =1,\ldots ,10$ and coefficient $b$, for (a): $a=10$, and (b): $a=20$.} 
 \label{fig5}
 \end{center}
\end{figure*}
\end{center}

A distributions of ${\rm {\mathcal B}}\left(\lambda _{i} \right)$ in function of the magnitude $M_{i} $ and coefficient $b$ are illustrated in \fref{fig5}. 

\begin{center}
\begin{figure*}[!h]
\begin{center}
\includegraphics[angle = 0,width=0.8\linewidth]{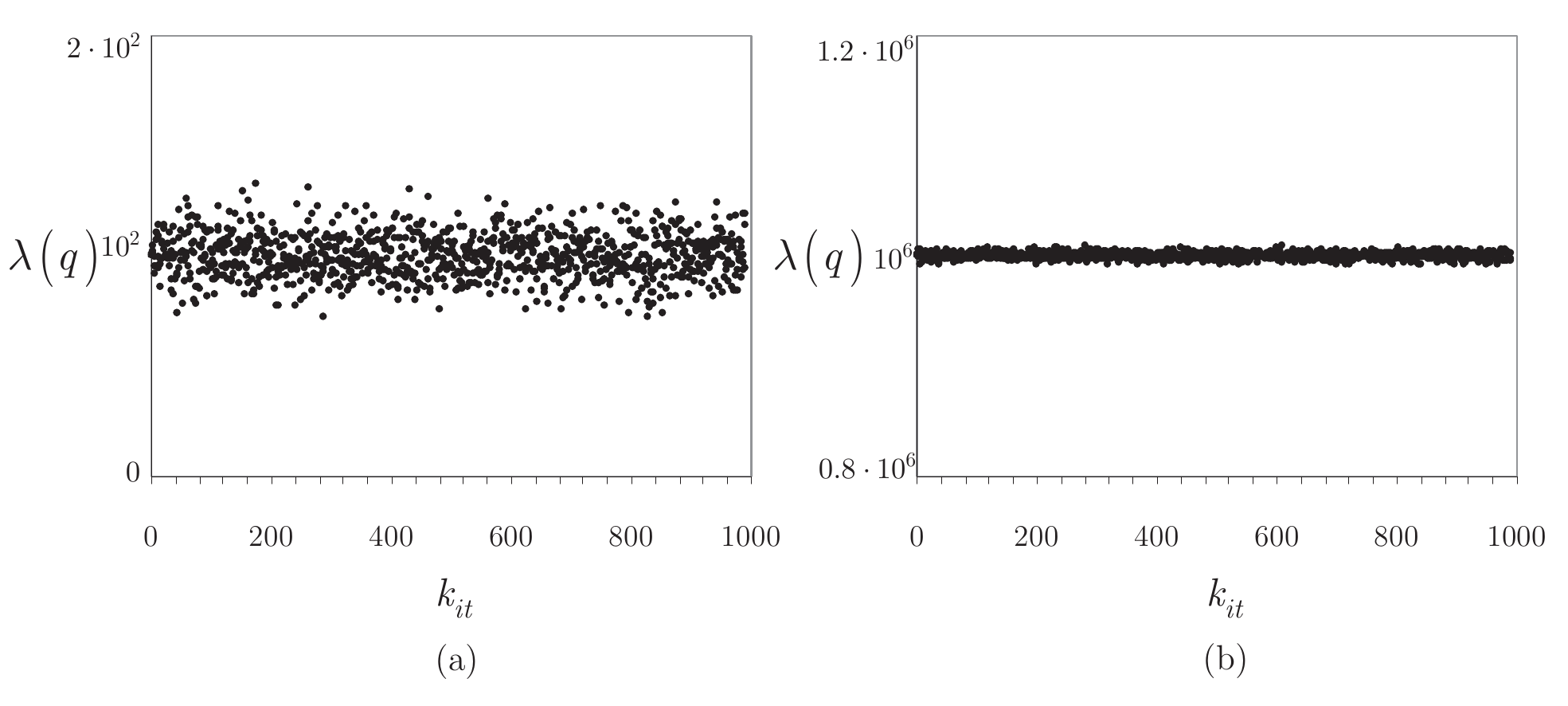}
\caption{Distribution of $\lambda \left(q\right)$ for $k_{it} $ iterations, $k_{it} =0,\ldots ,1000$, for (a): $\lambda \left(q\right)=10^{2} $, and (b): $\lambda \left(q\right)=10^{6} $.} 
 \label{fig6}
 \end{center}
\end{figure*}
\end{center}

The distributions of $\lambda \left(q\right)$ (see \eqref{ZEqnNum747242}) for $k_{it} $ iterations are depicted in \fref{fig6}. In \fref{fig6}(a), $\lambda \left(q\right)=10^{2} $, while \fref{fig6}(b) illustrated the distribution at $\lambda \left(q\right)=10^{6} $. As $\lambda \left(q\right)\to \infty $, the distributions of $\lambda \left(q\right)$ can be approximated by a ${\rm {\mathcal N}}\left(\lambda \left(q\right),\sigma _{{\rm {\mathcal N}}}^{2} \right)$, $\sigma _{{\rm {\mathcal N}}}^{2} =\lambda \left(q\right)$ Gaussian distribution.
  
\begin{center}
\begin{figure*}[!h]
\begin{center}
\includegraphics[angle = 0,width=0.8\linewidth]{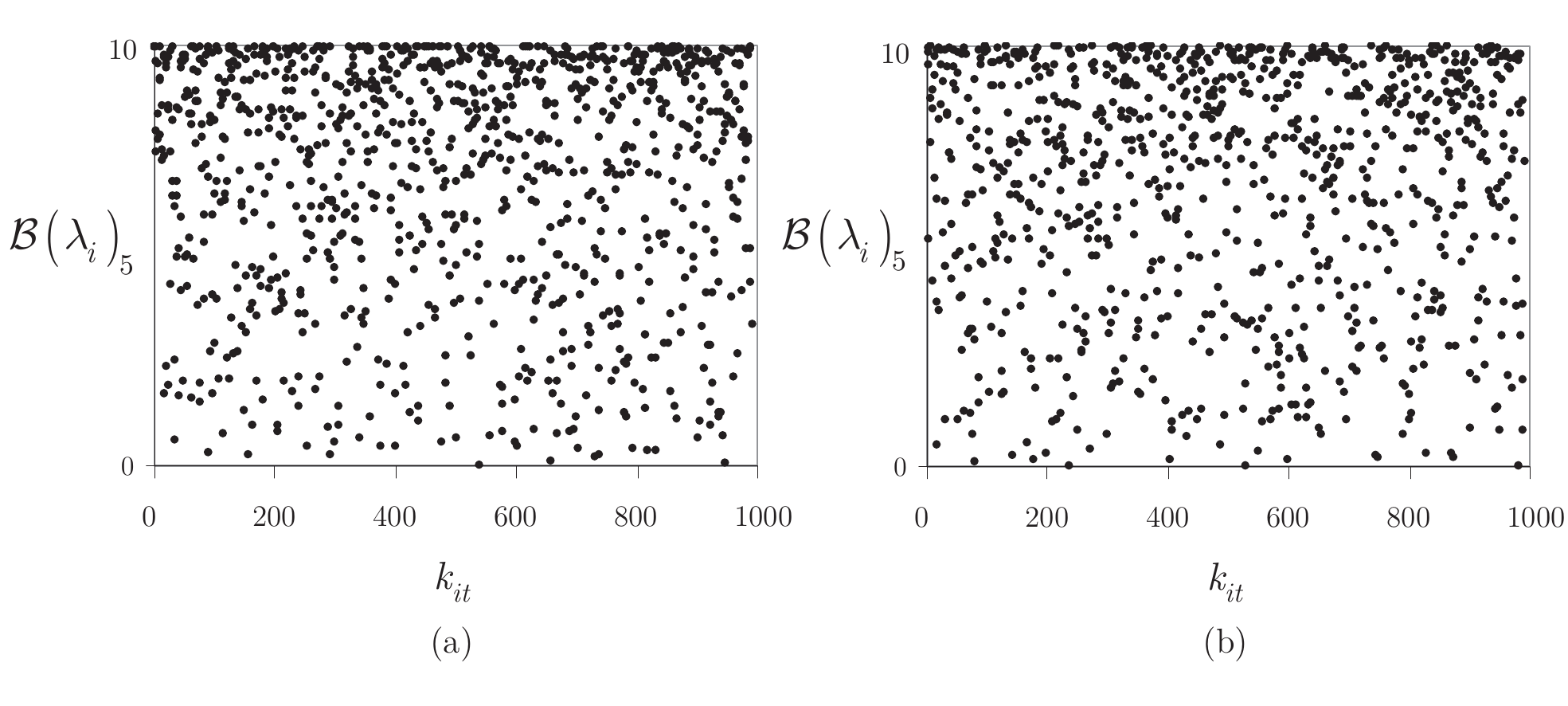}
\caption{The distributions of ${\rm {\mathcal B}}\left(\lambda _{i} \right)$ for $k_{it} $ iterations, $k_{it} =0,\ldots ,1000$, (a): $\lambda \left(q\right)=10^{2} $, and (b): $\lambda \left(q\right)=10^{6} $.} 
 \label{fig7}
 \end{center}
\end{figure*}
\end{center}

The associated distributions of ${\rm {\mathcal B}}\left(\lambda _{i} \right)$ for the values of $\lambda \left(q\right)$ are depicted in \fref{fig7}. The maximum value of ${\rm {\mathcal B}}\left(\lambda _{i} \right)$ is selected to ${\rm {\mathcal B}}\left(\lambda _{i} \right)\approx 10$ in each cases which values are picked up at $\lambda \left(q\right)$, where $\lambda \left(q\right)=10^{2} $ in \fref{fig7}(a), and $\lambda \left(q\right)=10^{6} $ in \fref{fig7}(b). The ${\rm {\mathcal B}}\left(\lambda _{i} \right)$ values approximates to a Gaussian distribution. The statistical distribution of ${\rm {\mathcal B}}\left(\lambda _{i} \right)$ is therefore constitutes a similar pattern for arbitrary $\lambda \left(q\right)$. 

\section{Conclusions}
\label{sec6}
We defined an optimization framework for the transmission and processing of quantum entanglement in the entangled network structure of the quantum Internet.
The proposed Poissonian entanglement optimization framework fuses the fundamental concepts of quantum Shannon theory with the theory of evolutionary algorithms and seismic wave propagations. Two objective functions are defined, with primary focus on the entanglement fidelity and secondary focus on the relative entropy of entanglement. As an additional objective function, the minimization of classical communications required by the entanglement optimization procedure is considered. The cost functions are defined to cover the physical attributes of entanglement transmission, purification, and storage in quantum memories. This method can be implemented with low complexity that allows a straightforward application in future quantum Internet and quantum networking scenarios. 


\section*{Acknowledgements}
This work was partially supported by the European Research Council through the Advanced Fellow Grant, in part by the Royal Society’s Wolfson Research Merit Award, in part by the Engineering and Physical Sciences Research Council under Grant EP/L018659/1, by the Hungarian Scientific Research Fund - OTKA K-112125, and by the National Research Development and Innovation Office of Hungary (Project No. 2017-1.2.1-NKP-2017-00001), and in part by the BME Artificial Intelligence FIKP grant of EMMI (BME FIKP-MI/SC).

\newpage
\appendix
\setcounter{table}{0}
\setcounter{figure}{0}
\setcounter{equation}{0}
\renewcommand{\thetable}{\Alph{section}.\arabic{table}}
\renewcommand{\thefigure}{\Alph{section}.\arabic{figure}}
\renewcommand{\theequation}{\Alph{section}.\arabic{equation}}

\section{Appendix}
\subsection{Definitions}
\label{appdef}
\subsubsection{Entanglement Fidelity}
Let
\begin{equation}
{\left| \beta _{00}  \right\rangle} ={\textstyle\frac{1}{\sqrt{2} }} \left({\left| 00 \right\rangle} +{\left| 11 \right\rangle} \right)
\end{equation}
be the target Bell state subject to be created at the end of the entanglement distribution procedure. The entanglement fidelity $F$ at an actually created noisy quantum system $\sigma $ is
\begin{equation}
F\left( \sigma  \right)=\langle  {{\beta }_{00}} |  \sigma |{{\beta }_{00}} \rangle ,
\end{equation}
where $F$ is a value between $0$ and $1$, $F=1$ for a perfect Bell state and $F<1$ for an imperfect state.
The fidelity for two pure quantum states is defined as 
\begin{equation} \label{eq1} 
F(| \varphi \rangle ,| \psi \rangle )=|\langle \varphi |\psi \rangle |^{2} . 
\end{equation} 
 The fidelity of quantum states can describe the relation of a pure channel input state $|\psi \rangle $ and the received mixed quantum system $\sigma =\sum _{i=0}^{n-1} p_{i} \rho _{i} =\sum _{i=0}^{n-1} p_{i} | \psi _{i} \rangle  \langle \psi _{i} | $ at the channel output as 
\begin{equation} \label{eq2} 
F(|\psi \rangle ,\sigma )=\langle \psi |\sigma |\psi \rangle =\sum^{n-1}_{i=0}{}p_i{|\langle \psi |{\psi }_i\rangle |}^2. 
\end{equation} 
 Fidelity can also be defined for mixed states $\sigma $ and $\rho $ 
\begin{equation} \label{eq3} 
F(\rho,\sigma)={(\text{Tr}(\sqrt{\sqrt{\sigma }\rho \sqrt{\sigma }}))}^2=\sum_i{}p_i{(\text{Tr}(\sqrt{\sqrt{{\sigma }_i}{\rho }_i\sqrt{{\sigma }_i}}))}^2. 
\end{equation} 

\subsubsection{Relative Entropy of Entanglement}
 By definition, the $E(\rho )$ relative entropy of entanglement function of a joint state $\rho $ of subsystems $A$ and $B$ is defined by the $D( \cdot \| \cdot )$ quantum relative entropy function, without loss of generality as 
\begin{equation} \label{eq4} 
E(\rho )=\mathop{{\min }}\limits_{\rho _{AB} } D( \rho \| \rho _{AB} )=\mathop{\min }\limits_{\rho _{AB} } \text{Tr}(\rho \log \rho )-\text{Tr}(\rho \log (\rho _{AB} )), 
\end{equation} 
 where $\rho _{AB} $ is the set of separable states $\rho _{AB}=\sum _{i=1}^{n} p_{i} \rho _{A,i} \otimes \rho _{B,i} $.

\subsection{Evaluation of Solutions}
\subsubsection{Fitness Function}
\label{app}
To evaluate the performance of the epicenters we utilize a mathematical apparatus based on the Pareto strength and fitness assignment \cite{ref9, ref10}. 

Let $\Pr \left({\rm {\mathcal E}}_{i} \right)$ be the probability of selection of an epicenter ${\rm {\mathcal E}}_{i} $, defined as
\begin{equation} \label{ZEqnNum262567} 
\Pr \left({\rm {\mathcal E}}_{i} \right)=\frac{\kappa \left({\rm {\mathcal E}}_{i} \right)}{\sum _{l\in K}\kappa \left({\rm {\mathcal E}}_{l} \right) } , 
\end{equation} 
where $\kappa \left({\rm {\mathcal E}}_{i} \right)$ is the sum of $d\left(\cdot \right)$ Euclidean distances between ${\rm {\mathcal E}}_{i} $ and the other epicenters, as
\begin{equation} \label{2)} 
\kappa \left({\rm {\mathcal E}}_{i} \right)=\sum _{l=1}^{K}d \left({\rm {\mathcal E}}_{i} ,{\rm {\mathcal E}}_{l} \right)=\sum _{l=1}^{K}\left\| {\rm {\mathcal E}}_{i} -{\rm {\mathcal E}}_{l} \right\|  , 
\end{equation} 
where $K$ is a set with cardinality 
\begin{equation} \label{3)} 
\left|K\right|=\sum _{i=1}^{\left|{\rm {\mathcal P}}\right|}D\left({\rm {\mathcal E}}_{i} \right)+ \sum _{i=1}^{N}\sum _{k=1}^{\dim \left({\rm {\mathcal E}}_{i} \right)}R\left(i,k\right)  ,                                        
\end{equation} 
where $D\left({\rm {\mathcal E}}_{i} \right)$ is given in \eqref{ZEqnNum754163}, and $l\in K$ refers to that the position of ${\rm {\mathcal E}}_{j} $ belongs to set $K$, and $\left|{\rm {\mathcal P}}\right|$ is the population size. Let ${\rm {\mathcal N}{\mathcal P}}$ refer to the non-dominated solution archive, and let
\begin{equation} \label{4)} 
\varphi _{i} ={\rm {\mathcal E}}_{i}  
\end{equation} 
refer to the selected epicenter, i.e, to an individual solution in ${\rm {\mathcal P}}$ or in ${\rm {\mathcal N}{\mathcal P}}$. 

Let $\Phi \left(\varphi _{i} \right)$ be a strength coefficient for solution $\varphi _{i} $, defined as
\begin{equation} \label{ZEqnNum866920} 
\Phi \left(\varphi _{i} \right)=\left|\varphi _{k} \in {\rm {\mathcal P}}\bigcup {\rm {\mathcal N}{\mathcal P}}\right|\left. \varphi _{k} \angle \varphi _{i} \right|,                                         
\end{equation} 
where $\angle $ refers to the Pareto dominance relation between $\varphi _{i} $ and $\varphi _{k} ={\rm {\mathcal E}}_{k} $. As follows, \eqref{ZEqnNum866920} depends on the number of individuals it dominates, by theory \cite{ref9, ref10}. 

By definition, a decision vector $\mathbf{A}$ dominates a vector $\mathbf{B}$, i.e., $\mathbf{B}\angle \mathbf{A}$, if 
\begin{equation} \label{6)} 
f_{i} \left(\mathbf{A}\right)\le f_{i} \left(\mathbf{B}\right) 
\end{equation} 
for $\forall i$, $i=1,\ldots ,m$ and for at least one $j$ with $i$, $j=1,\ldots ,n$,
\begin{equation} \label{7)} 
f_{j} \left(\mathbf{A}\right)\le f_{j} \left(\mathbf{B}\right),                                                 
\end{equation} 
where $f:{{\mathbb{R}}^{m}}\to {{\mathbb{R}}^{n}}$. The set of non-dominated decision vectors in ${\rm {\mathbb{R}}}^{n} $ is called a Pareto optimal set, while the image under $f$ in the solution space is called the Pareto front \cite{ref9, ref10}. In a multiobjective optimization the aim is to achieve the best Pareto front, by theory.    

Using \eqref{ZEqnNum866920}, let $\alpha \left(\varphi _{i} \right)$ be the raw fitness value of $\varphi _{i} $ evaluated by the $\Phi \left(\cdot \right)$ strength function (see \eqref{ZEqnNum866920}) of its dominators as
\begin{equation} \label{8)} 
\alpha \left(\varphi _{i} \right)=\sum _{\left(\varphi _{k} \in {\rm {\mathcal P}}\bigcup {\rm {\mathcal N}{\mathcal P}}\right)\wedge \left. \left(\varphi _{i} \angle \varphi _{k} \right)\right|}\Phi \left(\varphi _{k} \right) ,                                       
\end{equation} 
with an inverse distance function (referred to as the density value of $\varphi _{i} $), $\rho \left(\varphi _{i} \right)$ as
\begin{equation} \label{ZEqnNum791246} 
\rho \left(\varphi _{i} \right)=\frac{1}{d_{g} \left(\varphi _{i} \right)} ,                                                  
\end{equation} 
where $d_{g} \left(\varphi _{i} \right)$ is the distance from solution $\varphi _{i} $ to its $g^{th} $ nearest individual, where $g$ is initialized as the square root of the sample size $\left|{\rm {\mathcal P}}\bigcup {\rm {\mathcal N}{\mathcal P}}\right|$, by theory \cite{ref9, ref10}.

Using \eqref{ZEqnNum791246}, a for a random solution $r\left(\varphi _{i} \right)$ the $\tilde{f}\left(\cdot \right)$ fitness function of $\varphi _{i} $ is as
\begin{equation} \label{ZEqnNum608977} 
\tilde{f}\left(\varphi _{i} \right)=\alpha \left(\varphi _{i} \right)+\rho \left(\varphi _{i} \right).                                             
\end{equation} 
Then let $p$ refer to the number of selected $\varphi _{i} $ solutions in ${\rm {\mathcal P}}$. Using \eqref{ZEqnNum608977}, the selection probability of each solution is yielded as 
\begin{equation} \label{ZEqnNum469247} 
\Pr \left(\varphi _{i} \right)=\frac{\tilde{f}\left(\varphi _{i} \right)}{\sum _{r\in {\rm {\mathcal P}}}\tilde{f}\left(\varphi _{r} \right) } .                                                
\end{equation} 

\subsubsection{Constraints}

As a solution $\varphi _{i} $ does not satisfy the problem constraints $C_{1} $, $C_{2} $, $C_{3} $, a ${\rm H} ^{C_{z} } \left(\varphi _{i} \right)$, $z=1,2,3$  degrees of violation are defined for the constraints. 

For constraint $C_{1} $ (see \eqref{ZEqnNum992191}), the ${\rm H} ^{C_{1} } \left(\varphi _{i} \right)$ violation function \cite{ref9,ref10} is as 
\begin{equation} \label{ZEqnNum443575} 
H^{C_1}\left({\varphi }_i\right)=\left\{ \begin{array}{l}
\gamma -\zeta \left({\varphi }_i\right),\text{ if }\zeta \left({\varphi }_i\right)\le \gamma  \\ 
0,\text{otherwise}, \end{array}
\right. 
\end{equation} 
where
\begin{equation} \label{13)} 
\zeta \left(\varphi _{i} \right)=\sum _{i=1}^{N}{\rm {\mathcal F}}_{i} \left(\varphi _{i} \right) .                                                
\end{equation} 
For constraint $C_{2} $ (see \eqref{ZEqnNum104266}), the ${\rm H} ^{C_{2} } \left(\varphi _{i} \right)$ violation function is as follows
\begin{equation}  \label{ZEqnNum351305} 
{\rm H}^{C_2}\left({\varphi }_i\right)=\left\{ \begin{array}{l}
F_1\left({\varphi }_i\right)-\Lambda ,\text{ if }F_1\left({\varphi }_i\right)\ge \Lambda \\ 
0, \text{otherwise}, \end{array}
\right. 
\end{equation}
where
\begin{equation} \label{15)} 
F_{1} \left(\varphi _{i} \right)=\sum _{i=1}^{N}\sum _{i=1}^{T}f_{j} B_{F}^{j} \left(\varphi _{i} \right)  .                                            
\end{equation} 
For constraint $C_{3} $ (see \eqref{ZEqnNum901641}), the ${\rm H} ^{C_{3} } \left(\varphi _{i} \right)$ violation function is as 
\begin{equation} \label{ZEqnNum622567} 
{\rm H}^{C_3}\left({\varphi }_i\right)=\left\{ \begin{array}{l}
\nu \left({\varphi }_i\right)-\Pi ,\text{ if }\nu \left({\varphi }_i\right)\ge \Pi \\ 
0, \text{otherwise}, \end{array}
\right.
\end{equation}
where
\begin{equation} \label{17)} 
\nu \left(\varphi _{i} \right)=\sum _{j=1}^{N}\tau _{j} \left(\varphi _{i} \right) .                                                  
\end{equation} 
From \eqref{ZEqnNum443575}, \eqref{ZEqnNum351305} and \eqref{ZEqnNum622567} a penalty coefficient $\partial \left(\varphi _{i} \right)$ is defined as
\begin{equation} \label{18)} 
\partial \left(\varphi _{i} \right)=w_{1} {\rm H} ^{C_{1} } \left(\varphi _{i} \right)+w_{2} {\rm H} ^{C_{2} } \left(\varphi _{i} \right)+w_{3} {\rm H} ^{C_{3} } \left(\varphi _{i} \right),                             
\end{equation} 
where $w_{i} $-s are weighting coefficients \cite{ref9, ref10}. 
 
\subsubsection{Selection Condition}
Assuming that there are $\chi $ number of selected random solutions such that the selection probabilities are proportional to their fitness values. The selection of a solution $\varphi _{i} $ is as follows. 

First from the selected random solutions a mutant solution $\hbar _{i} $ is generated as
\begin{equation} \label{19)} 
\hbar _{i} =\varphi _{r_{a} } +\vartheta \left(\varphi _{r_{b} } -\varphi _{r_{c} } \right),                                                
\end{equation} 
where $r_{i} \in \left\{a,\ldots p\right\}$ are the random indexes, while $\vartheta >0$ is a coefficient.  

From the components of $\hbar _{i} $ a trial solution ${\rm T} _{i} $ is defined with a $j^{th} $ component ${\rm T} _{i}^{\left(j\right)} $ as
\begin{equation} \label{ZEqnNum568914} 
{\rm T}^{\left(j\right)}_i=\left\{ \begin{array}{l}
{\hslash }^{\left(j\right)}_i \text{ if } r\left(0,1\right)<P_{cross},\text{ or }j=r\left(i\right) \\ 
{\varphi }^{\left(j\right)}_i, \text{otherwise}, \end{array}
\right.
\end{equation} 
where $r\left(0,1\right)$ is a random number from the range $\left[0,1\right]$, $r\left(i\right)$ is a random integer within $\left(0,X\right]$ for each $i$, while $P_{cross} $ is the crossover probability ranged in $\left(0,1\right)$. 

Then the selection of the solution $\varphi _{i} $ using the trial solution ${\rm T} _{i} $ is as      
\begin{equation} \label{ZEqnNum985504} 
{\varphi }_i=\left\{ \begin{array}{l}
T_i,\text{ if }\tilde{f}\left(T_i\right)\le \tilde{f}\left({\varphi }_i\right) \\ 
{\varphi }_i, \text{otherwise}, \end{array}
\right. 
\end{equation} 
where function $\tilde{f}\left(\cdot \right)$ is given in \eqref{ZEqnNum608977}. 
 
\subsection{Sub-Procedure 1}
The Sub-procedure 1 of Algorithm 1 is as follows \cite{ref9, ref10}. 
\setcounter{algocf}{0}
\begin{subproc}
  \DontPrintSemicolon
\caption{}
Apply feasible space exploration \eqref{ZEqnNum379228} through the dimensions $L_{r}^{\dim _{k} \left({\rm {\mathcal E}}_{i} \right)} $ around $\dim _{k} \left({\rm {\mathcal E}}_{i} \right)$ of the epicenters. For $i=1,\ldots ,p$ obtain a ${\rm T} _{i} $ trial solution \eqref{ZEqnNum568914} for $\varphi _{i} $. Determine the best solution between  $\varphi _{i} $ and ${\rm T} _{i} $ via \eqref{ZEqnNum985504}. If $\tilde{f}\left({\rm T} _{i} \right)\le \tilde{f}\left(\varphi _{i} \right)$ and ${\rm T} _{i} $ is a non-dominated solution, then update ${\rm {\mathcal N}{\mathcal P}}$ with ${\rm T} _{i} $. Then, update ${\rm {\mathcal P}}$ with the best solution, and with other $p-1$ randomly selected solutions, $\varphi _{q} $, $q=1,\ldots ,p-1$, using the selection probability function \eqref{ZEqnNum262567} as $\Pr \left(\varphi _{q} \right)={\tilde{f}\left(\varphi _{q} \right) \mathord{\left/{\vphantom{\tilde{f}\left(\varphi _{q} \right) \sum _{i=1}^{p-1}\tilde{f}\left(\varphi _{i} \right) }}\right.\kern-\nulldelimiterspace} \sum _{i=1}^{p-1}\tilde{f}\left(\varphi _{i} \right) } $.
\end{subproc}

\subsection{Notations}
\setlength{\arrayrulewidth}{0.1mm}
\setlength{\tabcolsep}{5pt}
\renewcommand{\arraystretch}{1.5}
The notations of the manuscript are summarized in  \tref{tab1}.
\begin{center}
\begin{longtable}{||l|p{3.9in}||}
\caption{Summary of notations.}
\label{tab1}
\endfirsthead
\endhead
\hline
\textit{Notation} & \textit{Description} \\ \hline 
$l$  & Level of entanglement.  \\ \hline 
$F$ & Fidelity of entanglement.  \\ \hline 
$N$ & Number of nodes in the network. \\ \hline 
$T$ & Number of fidelity types $F_{j} $, $j=1,\ldots ,T$ of the entangled states. \\ \hline 
${\rm S}_{O} $ & Objective space. \\ \hline 
${\rm {\mathcal S}}_{F} $ & Feasible space. \\ \hline 
${\rm L}_{l} $ & An $l$-level entangled connection. For an ${\rm L}_{l} $ link, the hop-distance is $2^{l-1} $. \\ \hline 
$d\left(x,y\right)_{{\rm L}_{l} } $ & Hop-distance of an $l$-level entangled connection between nodes $x$ and $y$.  \\ \hline 
$E_{{\rm L}_{l} } \left(x,y\right)$ & entangled connection $E_{{\rm L}_{l} } \left(x,y\right)$ between nodes $x$ and $y$. \\ \hline 
$B_{F} \left(E_{{\rm L}_{l} } \left(x,y\right)\right)$ & Entanglement throughput of an ${\rm L}_{l} $-level entangled connection $E_{{\rm L}_{l} } \left(x,y\right)$ between nodes $\left(x,y\right)$. \\ \hline 
$B_{F}^{j} \left(x_{i} \right)$ & Number of incoming entangled states in an $i^{th} $ node $x_{i} $, with fidelity-type $j$, $i=1,\ldots ,N$. \\ \hline 
$\mathbf{X}$ & An $N\times T$ matrix, $\mathbf{X}={{\left( B_{F}^{j}\left( {{x}_{i}} \right) \right)}_{N\times T}}$, it describes the number of resource entangled states injected into the nodes from each fidelity-type in the network, $B_{F}^{j} \left(x_{i} \right)\ge 0$ for all $i$ and $j$.  \\ \hline 
${\rm {\mathcal F}}\left(x_{i} \right)$ & A primary objective function. It identifies the cumulative entanglement fidelity (a sum of entanglement fidelities in $x_{i} $) after an entanglement purification ${\rm P} \left(x_{i} \right)$ and an optimal quantum error correction ${\rm C}\left(x_{i} \right)$ in $x_{i} $. \\ \hline 
${\rm P} \left(x_{i} \right)$ & Entanglement purification in $x_{i} $. \\ \hline 
${\rm C}\left(x_{i} \right)$ & Optimal quantum error correction in $x_{i} $. \\ \hline 
$\left\langle B \right\rangle _{F}^{j}\left( {{x}_{i}} \right)$ & An initialization value for $B_{F}^{j} \left(x_{i} \right)$ in a particular node $x_{i} $. \\ \hline 
$\mathbb{E}\left( {{D}_{i}}\left( \mathbf{X} \right) \right)$ & A secondary objective function. It refers to the expected amount of cumulative relative entropy of entanglement (a sum of relative entropy of entanglement) in node $x_{i} $,  \\ \hline 
$w_{j} \left(x_{i} \right)$ & Quantum memory coefficient for the storage of entangled states from the $j^{th} $ fidelity type in a node $x_{i} $, evaluated as:\newline ${{w}_{j}}\left( {{x}_{i}} \right)={{\eta }_{j}}B_{F}^{j}\left( {{x}_{i}} \right)+{{\kappa }_{j}}\left\langle B \right\rangle _{F}^{j}\left( {{x}_{i}} \right)$,\newline where $\eta _{j} $ and $\kappa _{j} $ are coefficients to describe the storage characteristic of entangled states with the $j^{th} $ fidelity type.  \\ \hline 
$\tau _{j} \left(\mathbf{X}\right)$ & Differentiation of storage characteristic of entangled states from the $j^{th} $ fidelity type, defined as \newline $\tau _{j} \left(\mathbf{X}\right)=\sum _{i=1}^{N}\left(w_{j} \left(x_{i} \right)-\Omega \right)^{2}  ,$\newline where $\Omega ={\sum _{i=1}^{N}w_{j} \left(x_{i} \right)  \mathord{\left/{\vphantom{\sum _{i=1}^{N}w_{j} \left(x_{i} \right)  N}}\right.\kern-\nulldelimiterspace} N} $. \\ \hline 
$f_{C} \left({\rm P} \left(x_{i} \right)\right)$ & Cost of entanglement purification ${\rm P} \left(x_{i} \right)$ in $x_{i} $. \\ \hline 
$f_{C} \left({\rm C}\left(x_{i} \right)\right)$ & Cost of optimal quantum error correction ${\rm C}\left(x_{i} \right)$ in $x_{i} $. \\ \hline 
${\rm {\mathcal C}}\left(\mathbf{X}\right)$ & Total cost function, defined as\newline $\begin{array}{rcl}
   \mathcal{C}\left( \mathbf{X} \right)=&\sum\limits_{i=1}^{N}{{{f}_{C}}\left( P \left( {{x}_{i}} \right) \right)+{{f}_{C}}\left( \text{C}\left( {{x}_{i}} \right) \right)}  =\sum\limits_{i=1}^{N}{\sum\limits_{i=1}^{T}{{{f}_{j}}B_{F}^{j}\left( {{x}_{i}} \right),}}  
 \end{array}$\newline where $T$ is the number of fidelity types, $N$ is the number of nodes, $f_{j} $ is a total cost of purification and error correction associated to the $j^{th} $ fidelity type of entangled states. \\ \hline 
$f_{j} $ & Total cost of purification and error correction associated to the $j^{th} $ fidelity type of entanglement fidelity.  \\ \hline 
$F^{{\rm *}} $ & Critical fidelity coefficient. \\ \hline 
${\rm {\mathcal S}}_{low} $, ${\rm {\mathcal S}}_{high} $ & Sets with fidelity bounds ${\rm {\mathcal S}}_{low} \left(F\right)$ and ${\rm {\mathcal S}}_{high} \left(F\right)$ as\newline ${\rm {\mathcal S}}_{low} \left(F\right):\mathop{\max }\limits_{\forall i} F_{i} <F^{{*}} $,\newline and \newline ${\rm {\mathcal S}}_{high} \left(F\right):\mathop{\min }\limits_{\forall i} F_{i} \ge F^{{*}} $. \\ \hline 
$X_{{\rm {\mathcal S}}_{low} } $ & Set of nodes for which condition ${\rm {\mathcal S}}_{low} \left(F\right):\mathop{\max }\limits_{\forall i} F_{i} <F^{{*}} $ holds. \\ \hline 
$X_{{\rm {\mathcal S}}_{high} } $ & Set of nodes for which condition ${\rm {\mathcal S}}_{high} \left(F\right):\mathop{\min }\limits_{\forall i} F_{i} \ge F^{{*}} $ holds. \\ \hline 
${\rm {\mathcal S}}_{i} \left(\mathbf{X}\right)$ & Cost of quantum memory usage in node $x_{i} $, defined as \newline ${\rm {\mathcal S}}_{i} \left(\mathbf{X}\right)=\lambda \sum _{j=1}^{T}\alpha _{i} \frac{1}{\Upsilon _{i} }  B_{F}^{j} \left(x_{i} \right)$,\newline where $\lambda $ is a constant, $\alpha _{i} $ is a quality coefficient, while $\Upsilon _{i} $ is a capacity coefficient of the quantum memory. \\ \hline 
${\rm {\mathcal G}}\left(\mathbf{X}\right)$ & Main objective function, \newline $\mathcal{G}\left( \mathbf{X} \right)=\max \sum\limits_{i=1}^{N}{{{\mathcal{F}}_{i}}\left( \mathbf{X} \right)\mathbb{E}\left( {{D}_{i}}\left( \mathbf{X} \right) \right)} $. \\ \hline 
$F_{1} \left(N\right)$ & Minimization function for cost ${\rm {\mathcal C}}\left(\mathbf{X}\right)$. \\ \hline 
$F_{2} \left(N\right)$ & Minimization function for cost ${\rm {\mathcal S}}\left(\mathbf{X}\right)$. \\ \hline 
$C_{1} $, $C_{2} $, $C_{3} $ & Problem constraints. \\ \hline 
${\rm {\mathcal E}}$ & Epicenter, represents a solution in the feasible space. \\ \hline 
$L_{j} $ & A random location around epicenter ${\rm {\mathcal E}}$. \\ \hline 
$D\left({\rm {\mathcal E}}\right)$ & Dispersion coefficient of an epicenter ${\rm {\mathcal E}}$ (solution in the feasible space). It determines the number of affected $L_{j} $, $j=1,\ldots ,D\left({\rm {\mathcal E}}\right)$, locations (also represent solutions in the feasible space) around an epicenter ${\rm {\mathcal E}}$. \\ \hline 
${\rm {\mathcal P}}$ & Population ${\rm {\mathcal P}}$ (a set of possible solutions). \\ \hline 
$m$ & Control parameter. \\ \hline 
${\rm {\mathcal E}}_{i} $ & An $i^{th} $ individual (epicenter) from the $\left|{\rm {\mathcal P}}\right|$ individuals (epicenters) in the population ${\rm {\mathcal P}}$. \\ \hline 
$\tilde{f}\left(\cdot \right)$ & Fitness function. \\ \hline 
$\tilde{f}\left( \langle \mathcal{E} \rangle   \right)$ & A maximum objective value among the $\left|{\rm {\mathcal P}}\right|$ individuals. \\ \hline 
$\vartheta $ & A residual quantity. \\ \hline 
$f_{R} \left(\cdot \right)$ & Rounding function. \\ \hline 
$q$ & Total number of locations, $q=\sum _{i=1}^{\left|{\rm {\mathcal P}}\right|}D\left({\rm {\mathcal E}}_{i} \right) $. \\ \hline 
$\hat{D}\left({\rm {\mathcal E}}_{i} \right)$ & Upper bound on $D\left({\rm {\mathcal E}}_{i} \right)$ for a given epicenter ${\rm {\mathcal E}}_{i} $.  \\ \hline 
$d\left({\rm {\mathcal E}}_{i} ,l_{j} \right)$ & Euclidean distance $d\left({\rm {\mathcal E}}_{i} ,l_{j} \right)$ between an $i^{th} $ epicenter ${\rm {\mathcal E}}_{i} $ and the projection point $l_{j} $ of a $j^{th} $ location point $L_{j} $,$j=1,\ldots ,D\left({\rm {\mathcal E}}\right)$ on the ellipsoid around ${\rm {\mathcal E}}_{i} $. \\ \hline 
$\dim _{i} \left(\cdot \right)$ & An $i^{th} $ dimension of $l_{j} $. \\ \hline 
$P\left({\rm {\mathcal E}}_{i} ,L_{j} \right)$ & Seismic power $P\left({\rm {\mathcal E}}_{i} ,L_{j} \right)$ operator for an $i^{th} $ epicenter ${\rm {\mathcal E}}_{i} $. Measures the power in a $j^{th} $ location point $L_{j} $,$j=1,\ldots ,D\left({\rm {\mathcal E}}_{i} \right)$, as\newline $P\left({\rm {\mathcal E}}_{i} ,L_{j} \right)=\left(\frac{1}{d\left({\rm {\mathcal E}}_{i} ,l_{j} \right)} M\left({\rm {\mathcal E}}_{i} ,L_{j} \right)\right)^{b_{1} } b_{0} e^{\sigma _{\ln P\left({\rm {\mathcal E}}_{i} ,L_{j} \right)} } $,\newline where $b_{0} $ and $b_{1} $ are regression coefficients, $\sigma _{\ln P\left({\rm {\mathcal E}}_{j} \right)} $ is the standard deviation, while $M\left({\rm {\mathcal E}}_{i} ,L_{j} \right)$ is the seismic magnitude in a location $L_{j} $, while $l_{j} $ is the projection of $L_{j} $ onto the ellipsoid around ${\rm {\mathcal E}}_{i} $. \\ \hline 
$M\left({\rm {\mathcal E}}_{i} ,L_{j} \right)$ & Magnitude between epicenter ${\rm {\mathcal E}}_{i} $ and location $L_{j} $ is evaluated as\newline $M\left({\rm {\mathcal E}}_{i} ,L_{j} \right)=\left(P\left({\rm {\mathcal E}}_{i} ,L_{j} \right)\frac{1}{b_{0} e^{\sigma _{\ln P\left({\rm {\mathcal E}}_{i} ,L_{j} \right)} } } \right)^{\frac{1}{b_{1} } } d\left({\rm {\mathcal E}}_{i} ,l_{j} \right)$. \\ \hline 
$P^{{\rm *}} \left({\rm {\mathcal E}}_{i} \right)$ & Maximal seismic power for a given epicenter ${\rm {\mathcal E}}_{i} $. \\ \hline 
$C\left({\rm {\mathcal E}}_{i} \right)$ & Cumulative magnitude for an epicenter ${\rm {\mathcal E}}_{i} $. \\ \hline 
${\rm {\mathcal E}}'$ & Highest seismic power epicenter with magnitude $M\left({\rm {\mathcal E}}',L_{j}^{{\rm {\mathcal E}}'} \right)$. \\ \hline 
$\tilde{f}\left({\rm {\mathcal E}}'\right)$ & Minimum objective values among the $\left|{\rm {\mathcal P}}\right|$ epicenters. \\ \hline 
${\rm {\mathcal M}}$ & Control parameter, \newline ${\rm {\mathcal M}}=\sum _{i=1}^{\left|{\rm {\mathcal P}}\right|}M\left({\rm {\mathcal E}}_{i} ,L_{j}^{{\rm {\mathcal E}}_{i} } \right) $,\newline where $L_{j}^{{\rm {\mathcal E}}_{i} } $ provides the maximal seismic power for an $i^{th} $ epicenter ${\rm {\mathcal E}}_{i} $ \\ \hline 
$\Phi \left({\rm {\mathcal E}}_{i} ,{\rm {\mathcal R}}_{k} ,{\rm {\mathcal R}}_{l} \right)$ & Poisson range identifier function of ${\rm {\mathcal E}}_{i} $, where ${\rm {\mathcal R}}_{k} $ and ${\rm {\mathcal R}}_{l} $ are random reference points.  \\ \hline 
$c_{w} \left({\rm {\mathcal E}}_{i} ,{\rm {\mathcal R}}_{k} \right)$, $c_{w} \left({\rm {\mathcal R}}_{k} ,{\rm {\mathcal R}}_{l} \right)$ & Weighting coefficients between epicenters ${\rm {\mathcal E}}_{i} $ and ${\rm {\mathcal R}}_{k} $, and between ${\rm {\mathcal R}}_{k} $ and ${\rm {\mathcal R}}_{l} $. \\ \hline 
$\mathfrak{D}\left( {{\mathcal{E}}_{p}} \right)$ & Poissonian distance function $\mathfrak{D}\left( {{\mathcal{E}}_{p}} \right)$, where ${\rm {\mathcal E}}_{p} $ is a new solution.  \\ \hline 
$r\left({\rm {\mathcal E}}_{i} \right)$ & Radius around a current solution ${\rm {\mathcal E}}_{i} $, defined as\newline $r\left({\rm {\mathcal E}}_{i} \right)=\chi 10^{Q_{1} \left(2\tilde{{\rm {\mathcal M}}}\right)-Q_{2} } $,\newline where $\tilde{{M}}$ is the average magnitude \newline $\tilde{{M}}=\frac{1}{\left|{\rm {\mathcal P}}\right|} {\rm {\mathcal M}}=\frac{1}{\left|{\rm {\mathcal P}}\right|} \sum _{i=1}^{\left|{\rm {\mathcal P}}\right|}M\left({\rm {\mathcal E}}_{i} ,L_{j}^{{\rm {\mathcal E}}_{i} } \right) $,\newline while $Q_{1} $ and $Q_{2} $ are constants, while $\chi $ is a normalization term.  \\ \hline 
$\dim _{k} \left({\rm {\mathcal E}}_{i} \right)$ & Randomly selected $k^{th} $ dimension, $k=1,\ldots ,\dim \left({\rm {\mathcal E}}_{i} \right)$ of a current epicenter ${\rm {\mathcal E}}_{i} $, $i=1,\ldots ,\left|{\rm {\mathcal P}}\right|$. \\ \hline 
${\rm {\mathcal H}}\left(\dim _{k} \left({\rm {\mathcal E}}_{i} \right)\right)$ & Hypocentral, provides a random displacement of $\dim _{k} \left({\rm {\mathcal E}}_{i} \right)$ using $C\left({\rm {\mathcal E}}_{i} \right)$. \\ \hline 
$L_{r}^{\dim _{k} \left({\rm {\mathcal E}}_{i} \right)} $ & A random location in the $k^{th} $ dimension $L_{r}^{\dim _{k} \left({\rm {\mathcal E}}_{i} \right)} $ around $\dim _{k} \left({\rm {\mathcal E}}_{i} \right)$. \\ \hline 
${\rm N} \left(\cdot \right)$ & Normalization operator ${\rm N} \left(\cdot \right)$ of $L_{r}^{\dim _{k} \left({\rm {\mathcal E}}_{i} \right)} $. It keeps the new locations around $\dim _{k} \left({\rm {\mathcal E}}_{i} \right)$ in ${\rm {\mathcal S}}_{F} $, where $B_{low}^{k} $ and $B_{up}^{k} $ are lower and upper bounds on the boundaries of locations in a $k^{th} $ dimension.  \\ \hline 
$S$-metric & Hypervolume indicator. A quality measure for the solutions or a contribution of a single solution in a solution set. \\ \hline 
$S\left({\rm {\mathcal R}}\right)$ & $S$-metric for a solution set ${\rm {\mathcal R}}=\left\{r_{1} ,\ldots ,r_{n} \right\}$ is as\newline $S\left({\rm {\mathcal R}}\right)={\rm {\mathcal L}}\left(\bigcup _{r\in {\rm {\mathcal R}}}\left\{x_{ref} \angle x\angle \left. x\right|r\right\} \right),$\newline where ${\rm {\mathcal L}}$ is a Lebesgue measure, notation $b\angle a$ refers to that $a$ dominates $b$ (or $b$ is dominated by $a$), while $x_{ref} $ is a reference point dominated by all valid solutions in the solution set.  \\ \hline 
$f_{1} $, $f_{2} $ & Objective functions. \\ \hline 
${\rm {\mathcal C}}_{1} \left(x_{i} \right)$ & Cost results from the first-type classical communications related to a $x_{i} $. \\ \hline 
${\rm {\mathcal C}}_{2} \left(x_{i} \right)$ & Cost results from the second-type classical communications with respect to $x_{i} $. \\ \hline 
${\rm {\mathcal E}}^{{\rm *}} $ & Global optima. \\ \hline 
$m$  & Number of magnitude ranges. \\ \hline 
$n_{i} $ & Number of locations belonging to an $i^{th} $ magnitude range. \\ \hline 
${\rm {\mathcal B}}\left(n_{i} \right)$ & Power law distribution function for a log-scaled $n_{i} $,\newline ${\rm {\mathcal B}}\left(n_{i} \right):\log _{10} \left(n_{i} \right)=a-b\tilde{M}_{i} $,\newline where $\tilde{M}_{i} $ is a log scaled $M_{i} $, while $a$ and $b$ are constants.   \\ \hline 
$\tilde{n}_{i} $ & Poisson estimate of $n_{i} $, as \newline $\tilde{n}_{i} =\sigma _{i}^{2} =\lambda _{i} $,\newline where $\sigma _{i}^{2} $ is the observational variance, while $\lambda _{i} $ is the mean of a Poisson distribution. \\ \hline 
$\sigma _{q}^{2} $ & Estimated uncertainty, $\sigma _{q}^{2} =\lambda \left(q\right)=\sum _{i=1}^{m}f\left(\tilde{M}_{i} \right) $, where $f\left(\cdot \right)$ is a fitting function. \\ \hline 
$\lambda \left(q\right)$ & Mean total number, $\lambda \left(q\right)=\sum _{i=1}^{m}\lambda _{i}  \approx q$, where $\lambda _{i} $ is an $i^{th} $ component mean. \\ \hline 
${\rm {\mathcal B}}\left(\lambda _{i} \right)$ & Power law distribution function for $\lambda _{i} =\tilde{n}_{i} $. \\ \hline 
$k_{it} $ & Number of iterations. \\ \hline
\end{longtable}
\end{center}
\end{document}